# Embracing the chaos: Alloying adds stochasticity to twin embryo growth

Yang Hu [a], Vladyslav Turlo [b], Irene J. Beyerlein [c, d], Subhash Mahajan [e], Enrique J. Lavernia [a], Julie M. Schoenung [a], Timothy J. Rupert [a, b, *]

[a] Department of Materials Science and Engineering, University of California, Irvine, CA 92697, USA

[b] Department of Mechanical and Aerospace Engineering, University of California, Irvine, CA 92697, USA

[c] Mechanical Engineering Department, University of California, Santa Barbara, CA 93106, USA

[d] Materials Department, University of California, Santa Barbara, CA 93106, USA

[e] Department of Materials Science and Engineering, University of California, Davis, CA 95616, USA

* To whom correspondence should be addressed: trupert@uci.edu

**Abstract**

High-throughput atomistic simulations reveal the unique effect of solute atoms on twin variant selection in Mg-Al alloys. Twin embryo growth first undergoes a stochastic incubation stage when embryos choose which twin variant to adopt, and then a deterministic growth stage when embryos expand without changing the selected twin variant. An increase in Al composition promotes the stochastic incubation behavior on the atomic level due to nucleation and pinning of interfacial disconnections. At compositions above a critical value, disconnection pinning results in multiple twin variant selection.

Mg-rich alloys are attractive engineering materials because of their unique properties, with a key example being high specific strength as compared to other metals. For example, while Al or Fe alloys are widely used as structural materials in aerospace and automobile industry, replacing them with Mg alloys would significantly reduce the weight of aircraft and vehicles, as well as increase fuel efficiency [1,2]. However, Mg alloys tend to be brittle due to the limited number of easily-activated dislocation slip systems in hexagonal close-packed (hcp) materials [3,4]. Therefore, twinning becomes an important deformation mode during the plastic deformation of Mg alloys, and the ductility/formability of Mg alloys can be improved via the formation of twin meshes (defined as two or more intersecting arrays of twins). For example, Lentz et al. successfully constructed double twin meshes in Mg-Li alloys, with the resultant sample achieving large compression failure strains (>30%) [5]. A more recent work by Wang et al. also showed that higher ultimate tensile strength and two-fold increases in ductility was realized in pure Mg samples with gradient twin meshes consisting of intersecting $\{10\bar{1}2\}$ and $\{10\bar{1}1\}$ twins [6].

The formation of intersecting twin meshes requires the activation of multiple twin variants, which is usually explained based on high Schmid factor [7,8], local stress state [7,8], or minimization of the compatibility strain [9,10]. In contrast, the effect of alloying element additions is typically neglected. However, Cui et al. [11] reported deformed microstructures that contained significant differences between pure Mg and Mg-Al-Zn alloys, hinting that alloying can fundamentally change the twinning process. As shown in Fig. S1, the microstructure of the pre-compressed pure Mg, AZ31 (Mg-3.1Al-0.9Zn-0.45Mn, wt.%), and AZ91 (Mg-8.9Al-0.89Zn-0.42Mn, wt.%) samples show that there are more twins with smaller thicknesses in the alloy samples than the pure Mg sample that was deformed to the same extent. More importantly, there are increased numbers of twin variants being activated in the alloys than the pure Mg sample. The effect of solutes on twinning nucleation has been reported in Mg alloyed with rare earth elements [12,13], and reduced twin thickness has been observed in Mg alloys with solute segregation to the twin boundaries [14], but the effect of solute addition on twin variant selection has not been studied. Moreover, many studies on the growth mechanisms of mature twins in pure Mg and Mg alloys have been conducted, yet twin embryo growth mechanisms are largely neglected both experimentally and theoretically, despite the fact that the growth of twin embryos at early stages should influence the final twin morphology. For example, Luque et al. [15] and Spearot et al. [16] have described the growth of mature twins in deterministic ways using the idea that twin thickening occurs through the nucleation and propagation of disconnection loops or terraces. In these studies, twin boundary velocity (twin growth rate) was found to be a function of grain size, temperature, applied stress, and character of the disconnection loops nucleated on the twin plane (such as disconnection loop size, step height, etc.).

In the current study, molecular dynamics (MD) simulations of the early stages of twin embryo growth in Mg-Al alloys are used to isolate the effect of solute addition. A schematic of the triclinic simulation cell is shown in Fig. 1(a), with the X-axis set along the $[1\bar{2}10]$-direction and Y-axis set along the $[10\bar{1}1]$-direction. The top and bottom surfaces of the simulation cell are, therefore, parallel to the $(\bar{1}012)$ twin plane. The dimensions of the simulation box are approximately 5.1 × 77.2 × 77.0 nm$^3$ in the X, Y, and Z directions with periodic boundary conditions being applied to all directions. The sample contains ~1,300,000 atoms. One twin embryo with a length of 7.6 nm and a thickness of 4.1 nm is inserted at the center of a simulation box using the Eshelby method reported by Xu et al. [17]. The inset of Fig. 1(b) shows the morphology of the twin embryo, which is bound by twin boundaries (TBs), conjugate TBs (c-TBs, almost 90° to the TBs), basal-prismatic (BP), and prismatic-basal (PB) interfaces. Al atoms are then introduced by randomly replacing

Mg atoms, with the average concentration varying from 0-10 at.%. Note that the size of the simulation box for different alloy compositions was slightly different due to the change of equilibrium lattice constant as solute concentration was varied (see Fig. S3).

MD simulations are conducted using the Large-scale Atomic/Molecular Massively Parallel Simulator (LAMMPS) package [18] and an embedded-atom method (EAM) potential developed by Liu et al. [19]. Shear deformation is applied to the simulation box to drive embryo growth. First, a global shear stress of around 1.05 GPa is applied to the simulation cell using an NPT (isothermal-isobaric) ensemble at 1 K. The cell is held for 1 ps, by which time any fluctuations in the shear stress have become negligible. After this time, an NVT (isothermal-isochoric) ensemble is then used to fix the shear strain value that was responsible for the ~1.05 GPa stress (varies slightly with alloy composition) and relax the simulation cell. We note that the twin embryo wants to grow to relax the applied stress, which results in a drop in global shear stress as the twin fraction increases (Fig. S7). For Mg-Al alloys, sixty samples with the same global concentration but different distribution of solutes are used to enable a statistical analysis of twinning. For pure Mg samples, sixty samples are also used but with different initial atomic velocity distributions. Visualization of atomic configurations are performed using the OVITO software [20], while the Polyhedral Template Matching method [21] is used to characterize local crystalline structure and atomic orientation. Atoms in the matrix and twin are then differentiated from each other using this orientation information, with Mg atoms in the matrix colored red and those in the twinned region colored green. In all images, the Al solutes are colored dark purple. Details on the use of lattice orientation to differentiate twin from matrix are included in Section S2.

In our previous work on twin embryo growth in pure Mg [22], the lateral propagation of c-TBs was observed to be faster than the migration of TBs, which leads to TBs becoming the primary boundaries as the twin grows and matures. TB motion occurred by disconnection propagation, and a logarithmic relation between twin length and thickness was obtained. However, we observe here that these findings do not hold true and embryo growth in the alloy samples exhibit a very different growth behavior. Using Mg-7 at.% Al alloys as examples, Fig. 1(b) shows the evolution of twin embryos in two samples. We observe two different twinning behaviors, in which either the TB or c-TB becomes the primary boundary, leading to two different shapes. To follow the changes in shape, the twin length and thickness are measured as the position difference of the two TBs and c-TBs, respectively, with procedural details provided in Section S3. To allow for a better comparison of twin embryo growth among the sixty alloy samples, the twin length and thickness are normalized by the lattice constant, $a$, at a given concentration. The black dashed line in Fig. 1(b) is the 1:1 reference line, so the samples with horizontally- and vertically-grown twin embryos exhibit data points above or below the reference line, respectively. We note that the data generated near the end of the simulation, namely after the twin embryo reaches 22 vol.% of the simulation cell, is not used for any analysis due to the strong interaction with periodic images (see Section S4).

The first suspected cause for the two distinct evolutions in embryos is the variations in local stress between different random samples since twin embryos grow towards regions with large shear stress. However, we find that there is no discernible difference in the initial stress state in these samples (see Section S5 for additional discussion), while the deviation in twinning behavior occurs at very early growth stage. Thus, local stress variations are not the driving force for twin variant selection in our simulations.

To differentiate between the horizontally- and vertically-grown twin variants, it is convenient to use the aspect ratio ($AR$), which is defined here as a twin thickness divided by twin length.

Analyzing the two examples shown in Fig. 1(b), the atomic positions of the twin embryos at 0.5 vol.% and 22 vol.% overlap for the horizontally-grown embryo (Fig. 2(a)), while there are substantial deviations in atomic positions for the vertically-grown embryo (Fig. 2(b)), indicating rotation of basal planes during vertical twin embryo growth. An additional example of this basal plane rotation is shown in Fig. S10. Considering all simulation data, a linear dependence between basal plane rotation angle (defined as the misorientation between the basal plane and the YZ plane) and the natural logarithm of aspect ratio is revealed, as shown in Fig 2(c). The existence of this trend, as well as the analysis of experimental data (see Section S6), identifies that the horizontally-grown and vertically-grown twins belong to different co-zone twin variants, specifically the $(\bar{1}012)$ twin and $(10\bar{1}2)$ twin, corresponding to the negative and positive sign of $\ln(AR)$, respectively.

To understand the origins of the two states, we analyze the evolution of twin embryo shape using the variation of $\ln(AR)$ with the twin volume fraction ($V_t$), as shown in Fig. 2(d). Since the same initial embryo configuration is used for all samples, all of the curves start from the same negative value corresponding to the initial $(\bar{1}012)$ twin embryo. During the early growth stage, all of the curves strongly fluctuate up and down until they converge to either an increasing or decreasing trend, indicating the future twin variant. Eventually, as an indication of the two divergent behaviors, the value of $\ln(AR)$ becomes larger than 0 for some curves, indicating a transition to the $(10\bar{1}2)$ twin variant, while remaining below 0 for others. This analysis thus identifies two stages of twin embryo growth, a stochastic incubation stage and the deterministic growth stage, as denoted in Fig. 2(d).

The transition from one stage to another can be determined by tracking the slope $d\ln(AR)/dV_t$ for each curve (see Section S7), where the sign of the slope shows the intent to adopt one twin variant or another. The twin volume fraction corresponding to the transition point increases with an increase in solute composition (Fig. S15), making the stochastic stage more prominent. In turn, the probability of adopting the $(10\bar{1}2)$ twin variant increases with increasing solute composition, as demonstrated in Fig. 2(e). In this figure, $\ln(AR)$ is obtained for each sample at the end of the simulation run (22 vol.%), and plotted for the sixty samples at each alloy composition. The median values of $\ln(AR)$ for each composition are marked with black dots and shift closer to 0, meaning that there is more random selection of twin variant at higher compositions. Our simulations had to be halted once the embryo approached the cell boundaries. Occasionally, an embryo was growing in a way such that it would eventually choose the alternative $(10\bar{1}2)$ variant, but was stopped prematurely while still in the expected $(\bar{1}012)$ variant.

To extrapolate our MD results to larger length scales and better determine the chances to adopt another twin variant, we calculated the average slope $d\ln(AR)/dV_t$ in the range from 15 to 22 vol.% for each sample and present them in Fig. 2(f). Again, values below 0 would correspond to the expected $(\bar{1}012)$ twin, while values above 0 would indicate selection of the alternative $(10\bar{1}2)$ variant. Consistent with results from direct simulation, the extrapolation to larger simulation sizes suggest that alloying Mg with >5 at.% Al should promote the formation of multiple twin variants. The exact critical solute concentration beyond which there is possibility of forming multiple twin variants may be influenced by the applied stress and temperature or even the limitations of the interatomic potential to capture all nuances of the Mg-Al system, but the existence of this critical concentration itself is unexpected and profound. The fact that alloying can overcome the typical strong preference believed to be pre-determined by the relative orientation of the grain and applied stress has not been reported to date. Additional simulations were conducted to probe the effect of temperature, stress, initial embryo shape and alloying element type, with the results shown in Section S9. In all cases, we observe multiple twin variant selection, proving that this behavior is widespread.

Stochastic selection of multiple twin variants opens a pathway for the formation of twin meshes within individual grains. On the one hand, when the same macroscopic stresses are applied, a grain in an alloyed sample is more likely to develop multiple twin variants than a grain in a pure Mg sample, which could then result in intersecting twins which started as different twin embryos. On the other hand, we find via our simulations that the addition of more solutes also slows the growth of the twin embryos, potentially even restricting growth completely (Fig. S18 and Fig. S19). The critical solute composition for having multiple twin variants must be smaller than the one for completely restricting twin growth to have both (1) considerable twin embryo growth and (2) the random selection of twin variant. This requirement provides a possible criterion to choose the appropriate solute types.

To reveal the atomistic mechanisms behind random twin variant selection in the high-concentration alloy samples, we focus on the effect of solute atoms on the nucleation and propagation of twinning disconnections, since these two processes drive the twin embryo growth [22]. Twinning disconnections are defined as interfacial dislocations with step character and account for the formation and migration of TBs [23-25]. In our simulations, twinning disconnections can be nucleated in two ways, depending on the source of disconnection nucleation. As shown in Fig. 3(a), twinning disconnections are formed at BP/PB interfaces in pure Mg samples, which then migrate towards each other to advance the TB or c-TB. The black arrows mark the formation of new twinning disconnections, while solid black lines mark disconnection lines. Solute atoms have both supportive and restrictive impacts on embryo growth. On the one hand, solute atoms promote the nucleation of twinning disconnections at the TBs or c-TBs themselves [15,26,27]. In the alloy sample shown in Fig. 3(b), small disconnection loops (circled in black) have nucleated in the middle of the TB, indicating homogeneous nucleation of twinning disconnections without the assistance of the BP/PB interfaces. On the other hand, solutes exhibit a pinning effect and can reduce the velocity of twinning disconnection motion [15,26,27]. Disconnection lines in the alloy samples are more tortuous than the disconnection lines in pure Mg samples because of this pinning effect of solutes. While similar findings of solute effect on the nucleation and propagation of twinning disconnections during the growth of mature twins can be found in Refs. [15,26,27], our work further reveals that solute atoms also dramatically influence twin embryo growth. In Section S10, we show the velocity of TBs and c-TBs due to the interplay of disconnection nucleation and propagation for different alloys. In pure Mg samples, the c-TBs move much faster than TBs, corresponding to the formation of horizontally grown $(\bar{1}012)$ twins. As the solute concentration increases, the motion of c-TBs becomes slower and, at certain concentration level, the pinning effect of solutes is strong enough to induce the possibility of forming the vertically grown $(10\bar{1}2)$ twins. For a given individual embryo, the stochastic events of disconnection nucleation and pinning/unpinning lead to either higher TB velocity or higher c-TB velocity within an early stage of twin embryo growth, corresponding to the stochastic selection of different twin variants. This velocity difference will then maintain during the following deterministic growth stage, where there is embryo expansion without changing twin variant. BP/PB migration in a bicrystal geometry was also studied and confirmed that solutes slow the migration of these interfaces (see Section S11).

In summary, using atomistic simulations, we have discovered a process of random variant selection between $(\bar{1}012)$ and $(10\bar{1}2)$ co-zone twins in Mg-Al alloys after Al concentration reaches 5 at.% for the simulation geometry used here. The twin embryo growth can be separated into two stages: (1) the stochastic incubation stage when embryos choose which twin variant to eventually adopt, and (2) the deterministic growth stage when embryos expand without changing the selected twin variant. The stochastic twin variant selection is caused by the stochastic events of nucleation and pinning/unpinning of twinning disconnections by solute atoms. These findings reveal a previously unknown effect of alloying on twinning and thereby benefit continuously emerging strategies for improving Mg properties via alloy design. We also note that this stochastic twin variant selection could be used for the construction of twin meshes. Forming twin meshes requires twins that grow in different directions and intersect each other, which can be accomplished by random twin variant selection in Mg alloys even for the same macroscopic stress state.

**Acknowledgements**

This research was supported by the National Science Foundation through grants CMMI-1729829, CMMI-1729887, and CMMI-1723539.

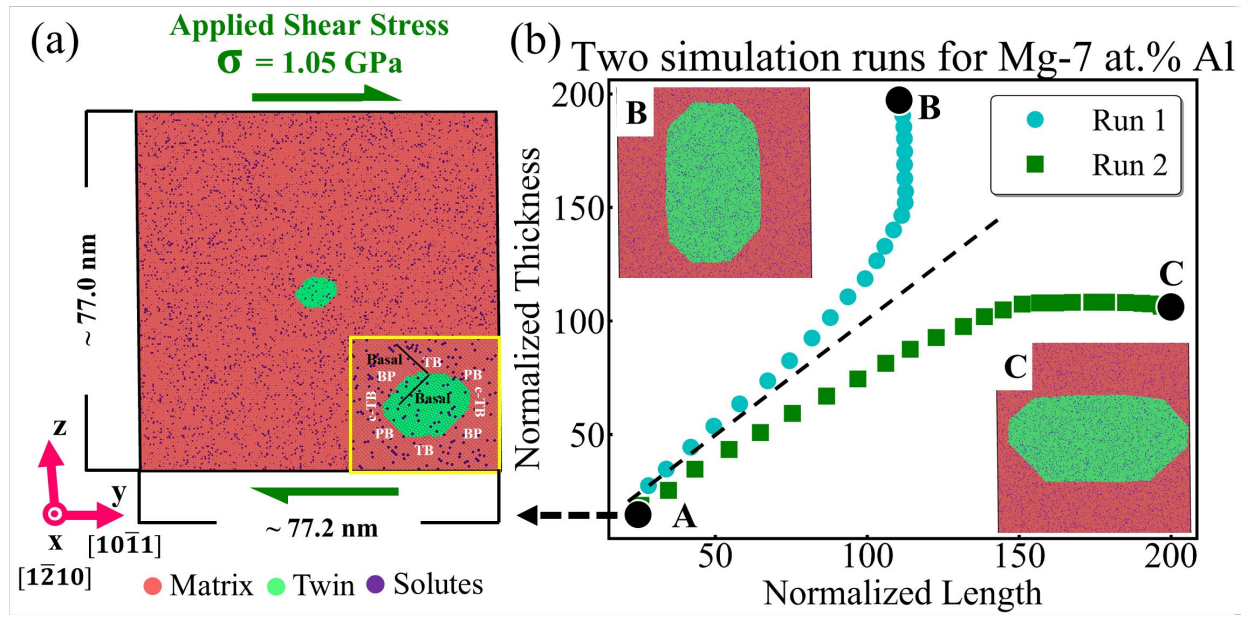

Fig. 1 (a) The atomic snapshot of the simulation box with the initial twin embryo shown. The inset of (a) shows an enlarged view of the twin embryo, and the basal planes in the matrix and the twin are marked using solid black lines. (b) The variation of normalized twin thickness with normalized twin length for two simulation runs of Mg-7 at.% Al. The insets of (b) show the final configurations of twin embryos in the two runs.

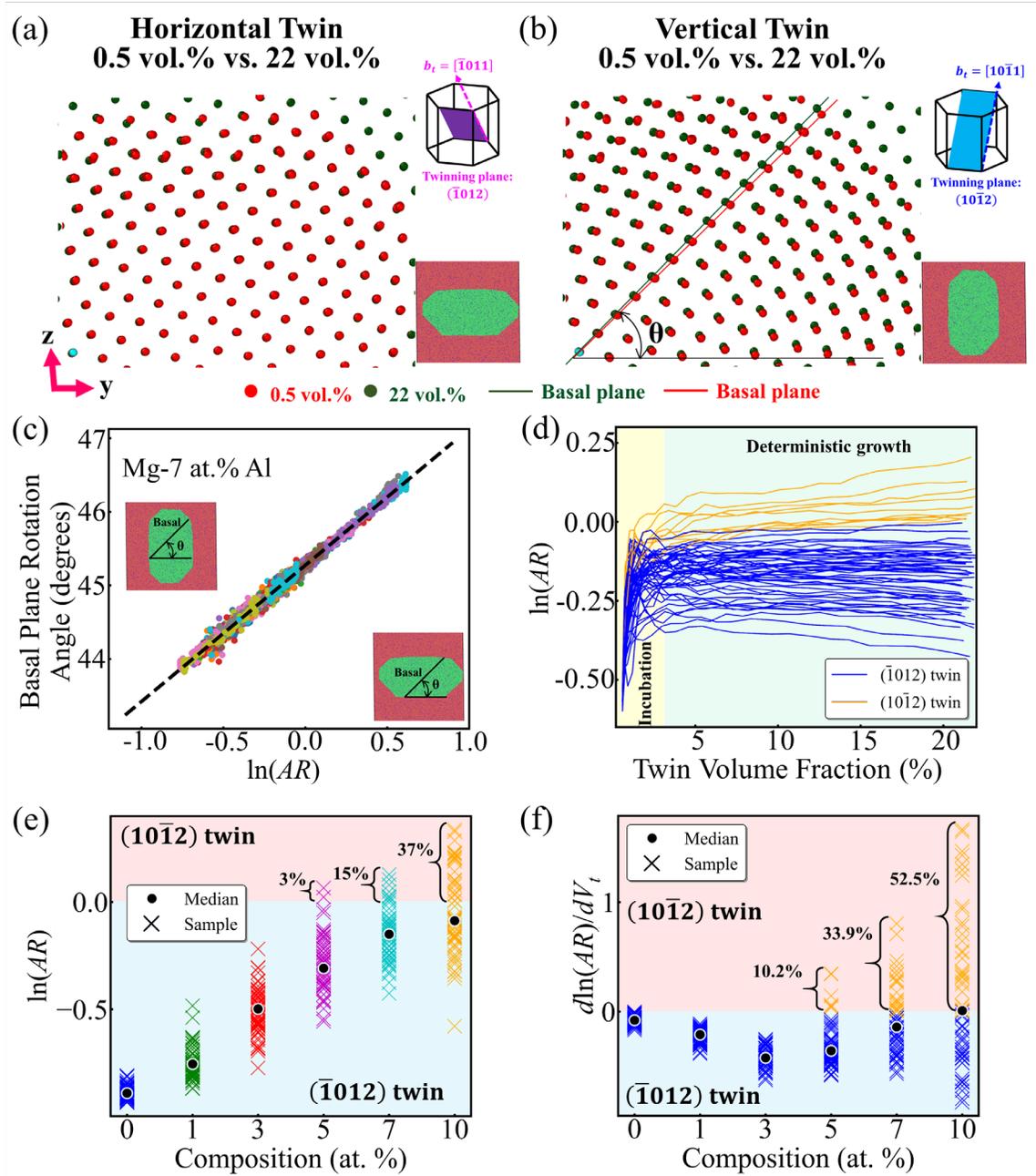

Fig. 2 The atomic positions of a twin embryo at the beginning (0.5 vol.%) and end (22 vol.%) of a simulation for (a) a sample which grows horizontally and has TBs as the primary boundaries and (b) a sample which grows vertically and has c-TBs as the primary boundaries. (c) Basal plane rotation angle versus the natural logarithm of aspect ratio for Mg-7 at.% Al samples, showing a linear relationship. (d) The variation of ln(AR) with twin volume fraction for Mg-7 at.% Al samples. (e) The natural logarithm of aspect ratio taken at the end of the simulations for pure Mg and Mg-Al alloy samples with different concentrations. The median of the data for each composition and the fraction of samples with ln(AR) larger than zero is also shown. (f) The average value of $d\ln(AR)/dV_t$ measured in the range from 15 to 22 vol.% in (e) is used to extrapolate to an infinitely large sample cell size.

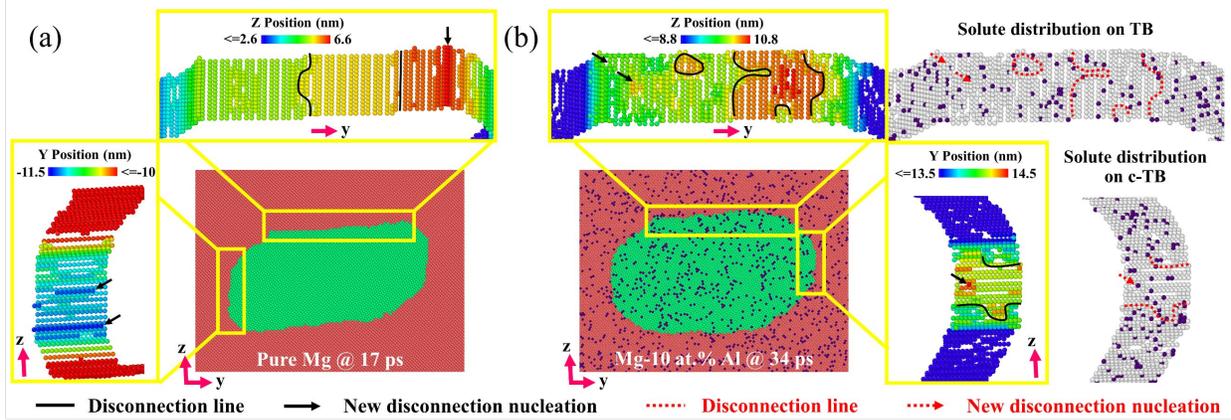

Fig. 3 Twinning disconnections formed on a TB and a c-TB in (a) pure Mg and (b) Mg-10 at.% Al samples. Atoms are colored using their Y positions in the enlarged views of c-TBs, while they are colored using their Z positions in the enlarged views of TBs. In (b), solute distribution on the TB and c-TB is also shown.

**Graphical abstract**

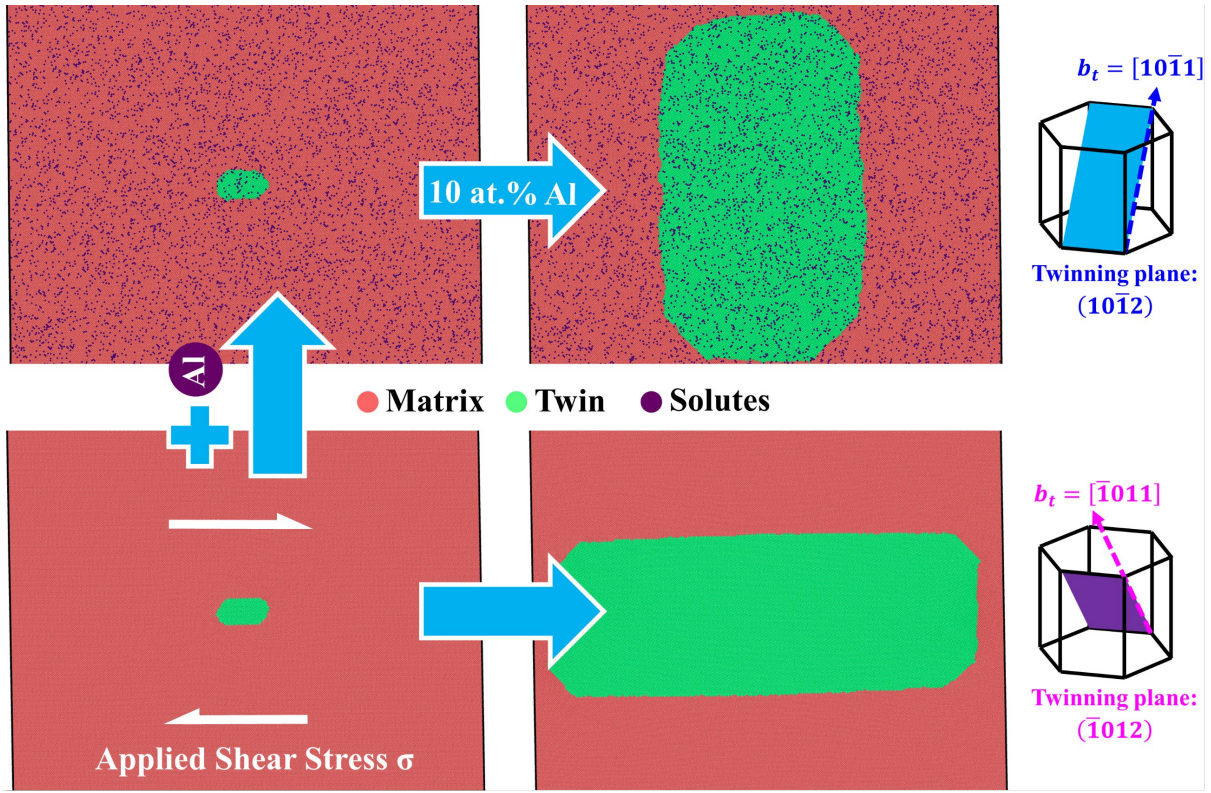

## S1. Twin morphology in pure Mg versus Mg-Al-Zn alloys

Cui et al. performed mechanical testing of pre-compressed Mg, AZ31 (Mg-3.1Al-0.9Zn-0.45Mn, wt.%), and AZ91 (Mg-8.9Al-0.89Zn-0.42Mn, wt.%) alloys to initiate further twinning or detwinning, with the microstructure of the pre-compressed state and after further deformation being tracked. Orientation maps of pure Mg versus two Mg-Al-Zn alloys pre-compressed at an engineering strain of 4% are presented in Fig. S1, with the images reproduced from Ref. [11]. All of these pre-compressed samples had similar grain sizes (~60 μm) and basal textures, and thus, the effect of grain size and initial texture on twinning and detwinning can be excluded. By comparing the microstructure shown in box A (pure Mg) and box E (AZ91), there are increased numbers of twin variants being activated in the alloys than the pure Mg sample.

## S2. Differentiating atoms in the twin and matrix using lattice orientation

The Polyhedral Template Matching method [22] implemented in the OVITO software [21] gives a point-to-point correspondence between the template structures and the structures to be analyzed, by minimizing the root-mean-square deviation of atom positions from the templates and simulated structures. The lattice orientation information can be encoded as an orientation quaternion, $q = q_w + q_x i + q_y j + q_z k$. Fig. S4(a) shows that component $q_w$ is the best to distinguish the twin embryo from the matrix, using a threshold value of $q_w = 0.8$. This criterion can be used for samples with different Al concentration and configurations taken at different times, as shown in Fig. S4(b).

## S3. Measuring the twin length and thickness

Twin length and thickness are measured as the position difference of two twin boundaries (TBs) and conjugate twin boundaries (c-TBs). To identify each boundary position, the simulation box is divided into 100 bins along the Y or Z-axis (displayed in Fig. S5), and then the number of atoms with $q_w > 0.8$ is counted in each bin and plotted versus the position of the bin. The peak in Fig. S5(c) then represents the twin embryo. Since the number of atoms with $q_w > 0.8$ in each bin changes with time as the twin grows, these values are normalized by the peak value in order to set up a consistent criterion for twin identification. Finally, the full width at half maximum values along the Y or Z-axis represent the positions of the TBs or c-TBs, respectively.

## S4. Defining the cutoff twin volume fraction

The twin volume fraction is plotted versus time in Fig. S6. Each curve starts with a tiny region where it is relatively flat, indicating a short time period for twin embryo growth to reach steady-state growth. Then there is an upward trend between the twin volume fraction and time until the primary boundary (TB or c-TB) reaches the edge of the simulation box, showing by another flat part of the curve. 22% is the largest twin volume fraction that can be chosen before strong interaction between twin embryos and their periodic images occur.

The appearance of plateaus near the end of simulations shown in Fig. 1(b) is an outcome of the interplay between the reduction of overall shear stress due to twin growth, and the influence coming from the periodic images of twin embryos (see Fig. S7). Shear stress is the activation force for embryo growth, it decreases as twin volume fraction increases. Therefore, the growth of embryos becomes very slow near the end of simulation. However, twin embryos tend to merge with their periodic images when the c-TBs/TBs approach the edge of the simulation box. So, for embryos growing in Y-direction, the twin length still increases, while for embryos growing in Z-direction, the twin thickness still increases. But adopting different twinning behavior is not determined by the strong interaction between the twin embryo and its periodic image near the end of the simulation.

## S5. 1D and 2D distribution of initial shear stress and hydrostatic stress

The spatial variation of initial shear stress and hydrostatic stress are plotted in Fig. S8 and Fig. S9, results for Mg-7 at.% Al are shown as examples. These two stresses are chosen because shear stress is the direct activation force for embryo growth, while hydrostatic stress shows how local lattices are distorted by solutes. The stresses at each Y position or Z position are averaged over samples showing the same twinning behavior, standard deviations are also calculated. In Fig. S8(a), the average shear stress decreases as the Y or Z position approaches the center of the simulation box, but suddenly jumps at certain positions, and then continue going down until it reaches the minimum. The two local peaks indicate the positions of TBs or c-TBs. Atoms in the vicinity of TBs and c-TBs are located in different structural environment than atoms in the bulk, and therefore leads to drastic changes of the local stress. Twinned regions show lower shear stresses caused by the twin shear transformation. Hydrostatic stresses fluctuate more than shear stresses and they are more uniformly distributed along the Y and Z-axis as shown in Fig. S8(b). Despite the different spatial variations of different stresses, the average stress distribution in samples adopting different twin variants are in fact very similar for both stresses. Though local fluctuations of the blue and orange curves slightly differ, the overall trend of the two curves is very similar and they almost overlap each other. Our results show that there is no obvious difference in the initial stress state in samples with different twin variants. Thus, stress is not the reason for twin variant selection in our simulations.

## S6. Determining the twin variants formed in simulation

To determine the misorientation between basal planes in different twin embryos, a basal plane rotation angle is first defined and calculated, which is the misorientation between the basal plane and the twin plane (small deviation of twin plane from the top and bottom surface of the simulation box is negligible, so we consider twin plane is parallel to the top and bottom surface of the simulation box, namely the YZ plane). The basal plane rotation angle is then plotted against the natural logarithm of the aspect ratio ($\ln(AR)$). The results for Mg-7 at.% Al are shown in Fig. S11(a) below. The MD data gives a linear trend between the basal plane rotation angle and $\ln(AR)$, $\alpha = k \cdot \ln(h/l) + b$, where $l$ and $h$ are twin length and thickness, $b$ is the basal plane rotation angle for a square shaped twin embryo (aspect ratio equal to 1). The basal plane rotation angle for two different twin embryos can be then expressed as follows,

$$\alpha_1 = k \cdot \ln\left(h_1/l_1\right) + b \tag{S1}$$

$$\alpha_2 = k \cdot \ln\left(h_2/l_2\right) + b \tag{S2}$$

The misorientation angle between basal planes in different twin embryos is calculated as $\alpha_1 - \alpha_2$. Note that we choose two embryos that satisfy the following condition for calculation,

$$\frac{h_1}{l_1} = \frac{l_2}{h_2} \tag{S3}$$

In this way we manage to relate the misorientation between basal planes to the aspect ratio of twin embryos, and a new parameter called shape factor can be defined. Shape factor = $|\ln(h/l)|$ is the absolute value of logarithm of aspect ratio, so the twin embryos with length and thickness satisfying Eqn. (S3) have the same shape factor. The basal plane rotation angle versus the shape factor is plotted in Fig. S11(b). It is clear that for embryos adopting different variants, the orientations of the basal planes are different. As the shape factor increases, meaning that either TBs or c-TBs become more dominant in the boundary of the twin embryo, the orientation of the basal planes in embryos adopting different variants becomes more different. Some orange circles (vertically-grown twin) are found among the blue circles (horizontally-grown twin), again indicating an incubation period of twin embryo growth. Fig. S11(c) shows the misorientation between the basal planes versus the shape factor. An experimental report of this misorientation is 4°~7° [10]. The linear trend predicted by MD data gives an estimation of the aspect ratio of 6.96 (corresponding shape factor is 1.94) at 7° misorientation, and the aspect ratio of 3.03 (corresponding shape factor is 1.11) at 4° misorientation. The mean value and standard deviation of the aspect ratios for selected twins from the work of Lou et al. [28] are calculated and shown in Fig. S12. Six twins are chosen for analysis, note that these are mature twins not twin embryos. The twin length and twin thickness is approximated as the length and width of the yellow rectangles, and the aspect ratio is then calculated as the length of the rectangle divided by width. A mean value of the aspect ratio is 6.80 (corresponding to a shape factor of 1.92), with upper limit of 9.67 and lower limit of 2.75 obtained. As this range overlaps with the range predicted by MD simulations, it gives us the confidence that we observe different twin variant in our simulations. Fig. S13 shows the deviation in basal plane orientation for co-zone twins from a crystallographic point of view.

## S7. Stochastic versus deterministic twin embryo growth

The boundary between the stochastic incubation period and deterministic growth stage is estimated as follows. The slope of each curve in Fig. S14(a), $d\ln(AR)/dV_t$, is first determined, then the $d\ln(AR)/dV_t$ of ±4 is marked in Fig. S14(b) due to the fact that each slope fluctuates within ±4 at the deterministic growth stage, then a vertical line is put where slope fluctuation starts to exceed the ±4 limit as the boundary between two growth stages. In the stochastic incubation period, twin embryos choose the variants to adopt and the choices are stochastic. While in the deterministic growth stage, twin embryos become larger with fixed twin variants. The same data in Fig. S14(a) is colored according to different twin variants and shown in Fig. 2(d). The blue and orange curves overlap in the incubation period, while there is a clear separation of them in the deterministic growth stage. Fig. S15 shows the slope of $\ln(AR)$ versus twin volume fraction for samples at different Al concentration. Alloys with higher solute concentration exhibit longer incubation period.

## S8. Raw data

The normalized twin length versus normalized twin thickness for all samples is shown in Fig. S16, with the 22 vol.% lines marked using grey dashed lines, which separate the curve of normalized twin length versus normalized twin thickness into two parts. Beyond ~22 vol.%, all curves start to plateau as twin embryo further expands.

## S9. The effect of stress, temperature, initial twin embryo shape, and alloying element type on stochastic twin variant selection.

To probe the effect of stress, temperature, initial twin embryo shape, and alloying element type on the formation of the vertically grown $(10\bar{1}2)$ twin, additional simulations were performed on Mg-10 at.% Al under different conditions. The twin variant selection data, presented as measurements of $\ln(AR)$ obtained at 22 vol.%, are shown in Fig. S17. We remind the reader that $\ln(AR)$ values above and below zero denote selection of different twin variants. Two additional temperatures (200 K and 300 K), two more stresses (1.28 GPa and 0.73 GPa), one different initial twin embryo shape ($h_0/l_0$ ~0.9), and one different alloying element (Y) were studied. For the Mg-10 at.% Al alloy deformed at 200 K and 300 K, and the Mg-10 at.% Y alloy, ten simulation runs were conducted. For other conditions, sixty samples were still used. An EAM potential developed by Sheng at el. [1] was used to simulate the Mg-Y alloy system.

Fig. S17 shows that different simulation conditions do not change the ability of Mg-10 at.% Al alloy to form the other twin variant. The mean value of $\ln(AR)$ approaches zero, with individual examples existing above and below this threshold, denoting formation of different twin variants. However, for samples deformed at higher stresses and temperatures, the spread in the data reduces somewhat and the fraction of $(10\bar{1}2)$ twins is slightly lower compared to simulations conducted at 1 K and 1 GPa. The possibility of forming the second twin variant is higher when there is strong pinning on the c-TB motion (see Section S10), yet higher stresses promote the unpinning of c-TBs from solutes and result in a faster motion. Therefore, we expect that the critical solute composition to trigger the possibility of forming the other twin variant will be higher at higher stresses. Temperature has a similar effect as stress, with increase temperature leading to faster motion of the c-TBs. While a thorough investigation of these variables is beyond the scope of this work, one can infer that the critical composition needed to observe multiple twin variant selection will increase as temperature and stress are increased. We do also note that temperature and stress are usually connected to one another, with, for example, materials plastically deforming at lower stresses as temperature is increased.

The formation of both $(\bar{1}012)$ and $(10\bar{1}2)$ twins is also observed in the Mg-10 at.% Y alloy. Y and Al have very different properties, as Al has a smaller atomic radius than Mg while Y has a larger atomic radius than Mg. They also have different crystal structure, bulk modulus, electronegativity, etc. in their monatomic forms. However, they both show the ability to induce the formation of the second twin variant. Yi and Falk [2] have discussed the effect of Al and Y on the motion of coherent twin boundaries, where these authors reported that both solutes promote the nucleation of twin loops while slow down the expansion of such twin loops. They also suggest that Al more strongly influences these factors than Y. Therefore, our hypothesis is that as long as the added solute atoms exhibit these effects on disconnection nucleation and propagation, it is possible to form the second twin variant.

**S10. Comparison of TB velocity and c-TB velocity in all alloys**

The velocity of TBs and c-TBs is calculated as the slope of the curves for twin length versus time and twin thickness versus time. As shown in Fig. S20, the two velocities (orange circles for TB and blue circles for c-TB) are plotted versus twin volume fraction for different alloys. In pure Mg samples, the formation of the horizontally grown twin corresponds to the faster motion of c-TBs than TBs, and the deviation in velocity occurs at a very early embryo growth stage. In alloy samples, those with low solute concentration behave similarly to pure Mg samples. However, after the solute concentration reaches a certain level, the ability to form the vertically grown twin appears and the velocity curves overlap. In some cases, the TB grows faster. In others, the c-TB grows faster. In Fig. S20, as the solute concentration increases the overall effect on disconnection nucleation and propagation leads to much slower c-TB motion, which eventually provides the possibility of forming the vertically grown twin. At higher solute concentration, the probability of vertically grown twins is higher due to the higher level of randomness (local composition fluctuation, etc.) in the sample introduced by higher amount of solutes. This induces more stochastic behavior in twin variant selection. Although TB motion is also inhibited by solutes to some extent, the pinning effect is much stronger on the c-TB, which was the boundary that moved faster in the pure Mg and lead to horizontal twin growth.

In Fig. S21, the comparison of TB velocity and c-TB velocity in individual samples of pure Mg and Mg-10 at.% Al alloy is shown. In pure Mg samples, the net effect of disconnection nucleation and disconnection propagation leads to a clear separation of the TB and c-TB velocity at early embryo growth stage. As twin embryos grow larger, this difference in boundary velocity causes a clear difference in twin length and twin thickness, which in turn leads to different stress fields around the TB and c-TB. In pure Mg samples, specifically, the stress field in front of the c-TB becomes more intense and can further promote boundary motion. Therefore, the boundary that moves faster initially will keep moving faster, corresponding to no change of the twin variant. For the alloy samples, due to the existence of solute atoms, the velocity curves fluctuate more severely and both boundaries move slower than in the pure Mg samples. Yet from Fig. S21 one can still find either faster TB motion or faster c-TB motion at the early embryo growth stage for a given sample, with the choice of boundary that moves faster being a stochastic result of the interplay of the events of disconnection nucleation and pinning/unpinning of disconnection propagation. Fig. S22 shows the local shear stress in a sample of pure Mg and Mg-10 at.% Al alloy. The two twin embryos both have a larger length than thickness, and two low shear stress regions on top/bottom of the TBs are marked using solid black lines. Although the existence of solute atoms complicates the shear stress state in alloys, the low shear stress regions can still be found.

**S11. The motion of basal-prismatic and prismatic-basal (BP/PB) interfaces**

Due to the small, nanoscale dimensions of the BP/PB interfaces, it is difficult to identify and decouple the effect of solute atoms on their motion. Thus, we performed additional simulations of BP/PB migration in a bicrystal geometry and confirmed that solutes also slow the migration of these interfaces. A schematic of the simulation box is shown in Fig. S23(a), where Mg atoms with a hexagonal close-packed (hcp) structural environment are colored blue and Mg atoms in a defect environment are colored white. The X-axis of the simulation box is parallel to the [$\bar{1}010$]-direction, the Y-axis is parallel to the [$1\bar{2}10$]-direction and the Z-axis is parallel to the [0001]-direction. The dimensions of the simulation box are around 22.6 × 22.9 × 92.6 nm$^3$ in the X, Y, and Z directions, respectively, and the sample contains ~2,130,000 atoms. The region in the middle of the box bound by two BP/PB interfaces is the prismatic lattice region. Periodic boundary conditions are applied to all directions. 0-10 at.% of Al solutes are introduced into the simulation box to randomly replace Mg atoms. To initiate BP/PB migration, a constant tensile/compressive stress of around 2 GPa is applied to the simulation box, with the direction parallel to the Z direction. Atoms are then relaxed under an NPT (i.e. isothermal-isobaric) ensemble at 1 K using Nose-Hoover thermostat, with the temperature adjusted every 100 time steps with one integration step of 0.1 fs. Smaller time step is used for better capturing the nanoscale phenomena occurred at the interface. By applying tensile stress, the prismatic lattice region expands, while the prismatic lattice region shrinks if compressive stress is applied. 2 GPa of stress is chosen because the Schmid factor of ($\bar{1}012$) and ($10\bar{1}2$) twins is about 0.5, therefore the BP/PB interface motion is comparable to that in the simulations on twin embryo growth. The thickness of the prismatic lattice region is normalized by the lattice constant at each concentration and plotted against time, which is shown in Fig. S23(b) and Fig. S23(c). The black dashed line shows after what time the stress fluctuation around 2 GPa becomes moderate. For the case of the shrinkage of the prismatic lattice region, all the curves for alloy samples are above the curve for pure Mg sample. While for the case of the expansion of the prismatic lattice region, all the curves for alloy samples are below the curve for pure Mg sample. The fact that the BP/PB interfaces move the fastest in the pure Mg samples demonstrate the pinning effect of solutes on BP/PB interface motion. The atomic snapshots of upper and lower BP/PB interfaces at the same time (6 ps) for pure Mg, Mg-3 at.% Al and Mg-10 at.% Al can be found in Fig. S24, the views from two perspectives are shown. Atoms are colored using their Z positions, with red color for atoms on the top and dark blue color for atoms at the bottom. The interfaces in alloy samples are rougher, indicating more disconnection nucleation in the alloy sample than pure Mg sample. But an overall effect of solute atoms is reducing the velocity of BP/PB interfaces.

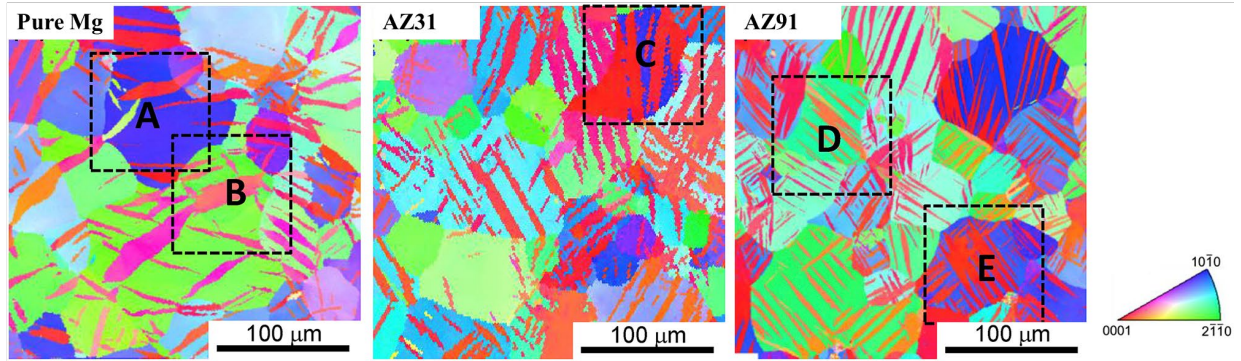

Fig. S1 Grain orientation maps of pure Mg, AZ31 alloy, and AZ91 alloy compressed to an engineering strain of 4% under a strain rate of 0.01 s$^{-1}$. Images are reproduced with permission from Ref. [11]. As can be seen by comparing the microstructure shown in box A (pure Mg) and box E (AZ91), there are increased numbers of twin variants being activated in the alloys than the pure Mg sample.

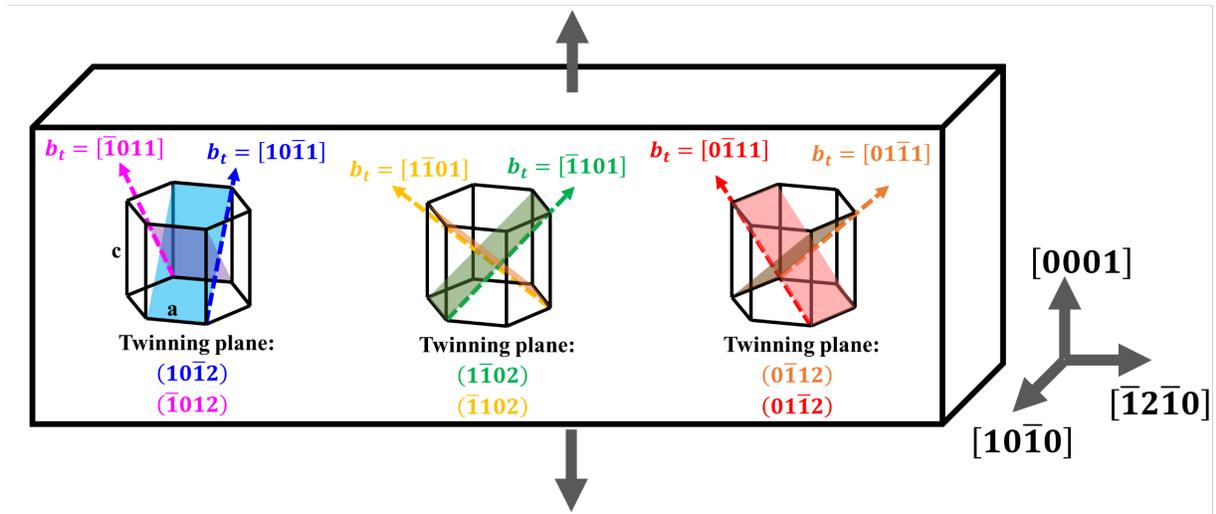

Fig. S2 Schematic of six twin variants for $\{10\bar{1}2\}$ tension twin. These six twin variants can be initiated by stretching the [0001]-axis (c-axis). Hexagons are used to show the unit cell of Mg, with the twinning planes shown using parallelograms. Arrows mark the twinning direction for each twin variant.

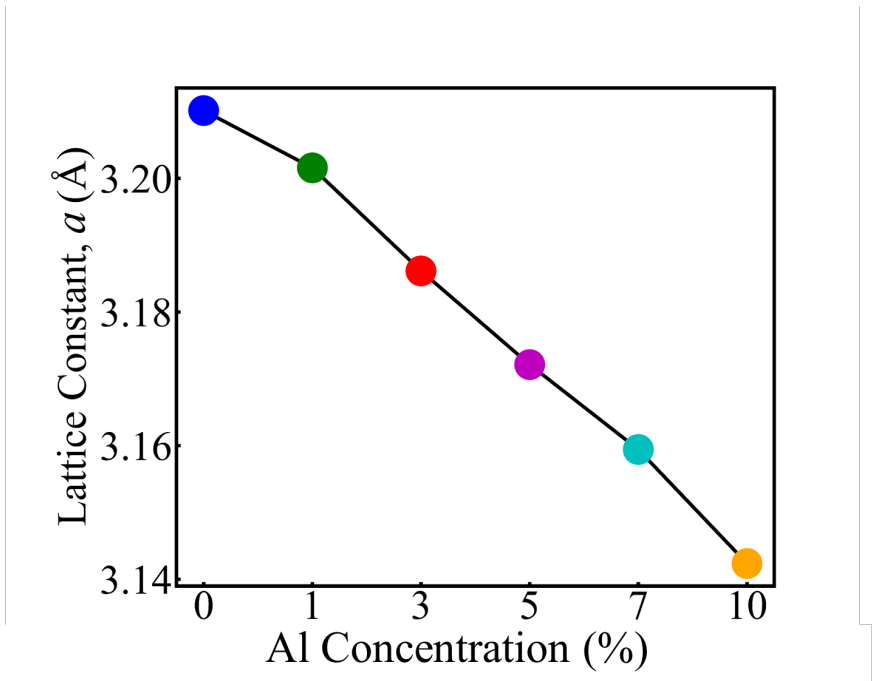

Fig. S3 The variation in lattice constant, *a*, with Al concentration.

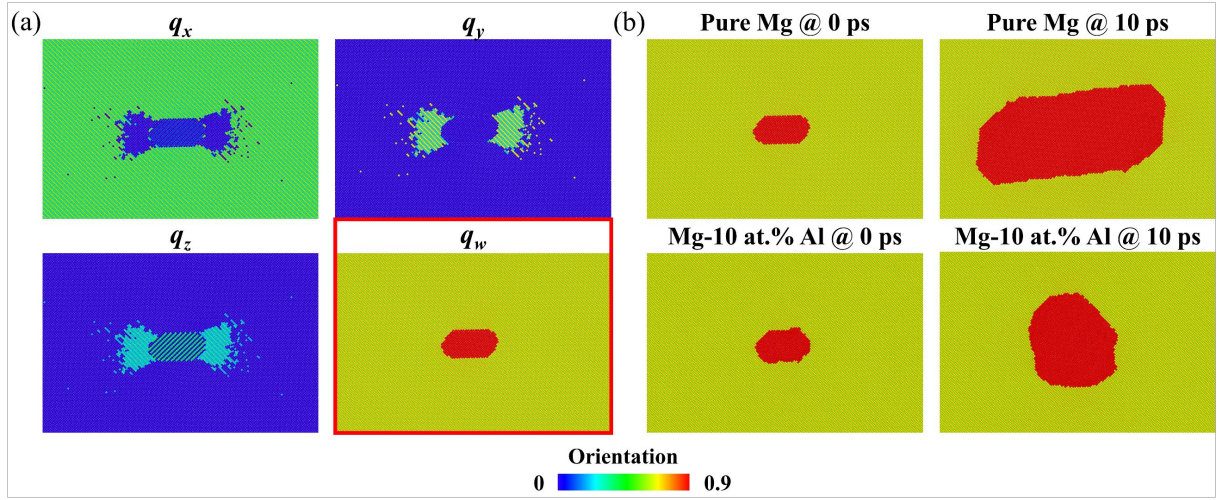

Fig. S4 (a) The local orientation information associated with each atom in the MD samples obtained by Polyhedral Template Matching. The twinned region can be correctly selected by $q_w$ with a value larger than 0.8, and this criterion is later used in determining the boundaries of the twin embryo. (b) The twinned region selected by the criterion, $q_w > 0.8$, in samples of pure Mg and Mg-10 at.% Al and at 0 ps and 10 ps, respectively.

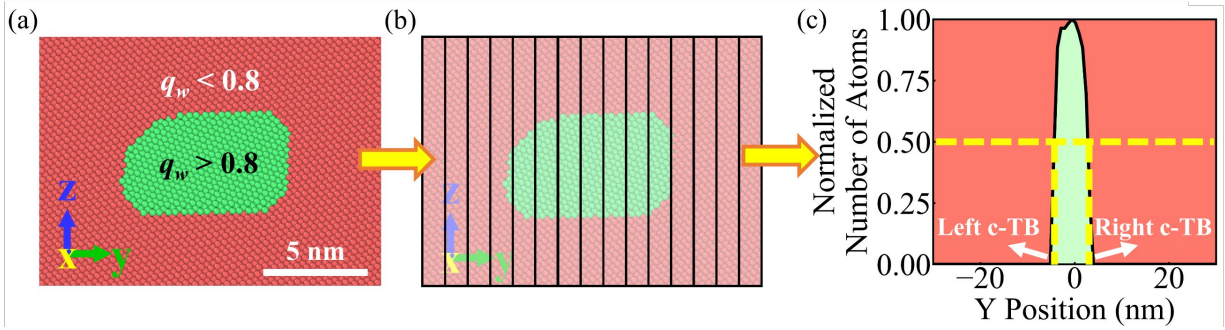

Fig. S5 Determining the positions of twin embryo boundaries. (a) The twinned region can be accurately detected using lattice orientation information. The twin embryo is colored green to contrast with the red matrix. (b) The simulation box is then divided into several bins, and the number of atoms with $q_w$ larger than 0.8 in each bin is counted. (c) Next, these numbers are normalized by their maximum value and plotted versus the positions of the bins, where the peak represents the twinned region while the locations of boundaries are chosen as the two positions with Y-axis value equal to 0.5.

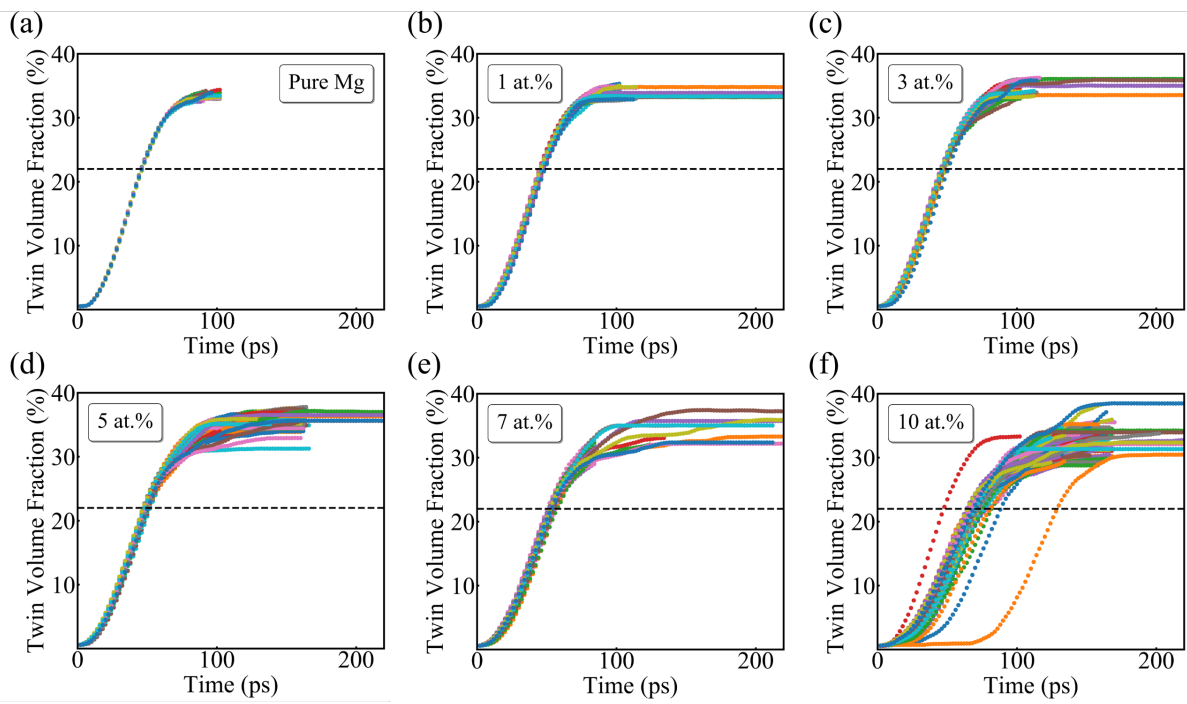

Fig. S6 Raw data of twin volume fraction versus time for (a) pure Mg, (b) Mg-1 at.% Al, (c) Mg-3 at.% Al, (d) Mg-5 at.% Al, (e) Mg-7 at.% Al, and (f) Mg-10 at.% Al. Twin volume fraction of 22% is marked using black dashed lines.

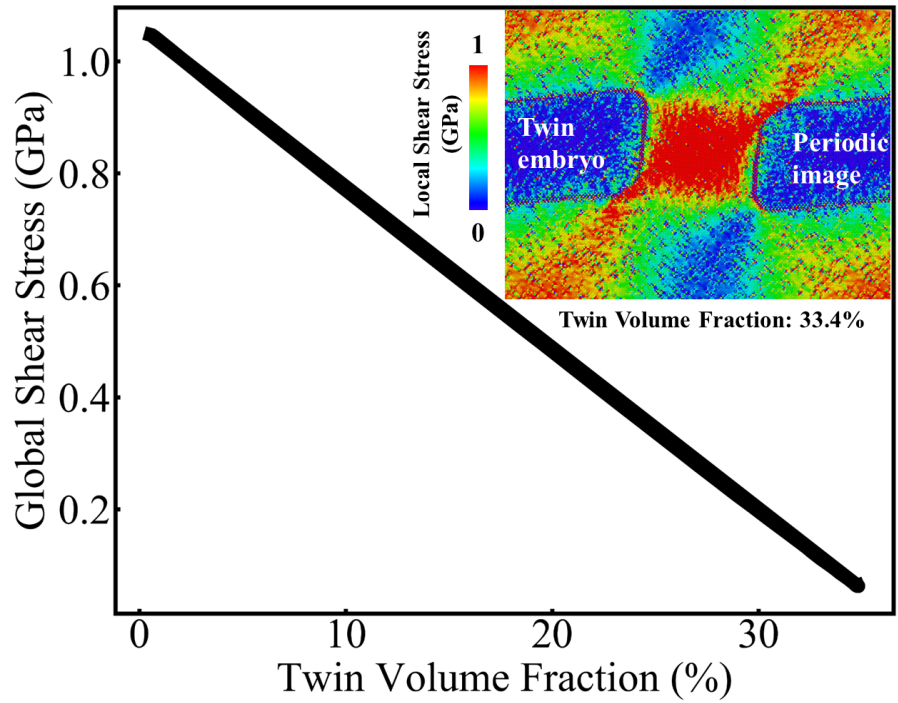

Fig. S7 The variation of the global shear stress with the twin volume fraction for one sample of Mg-1 at.% Al. The inset shows the distribution of local shear stresses around the twin embryo and its periodic image near the end of simulation.

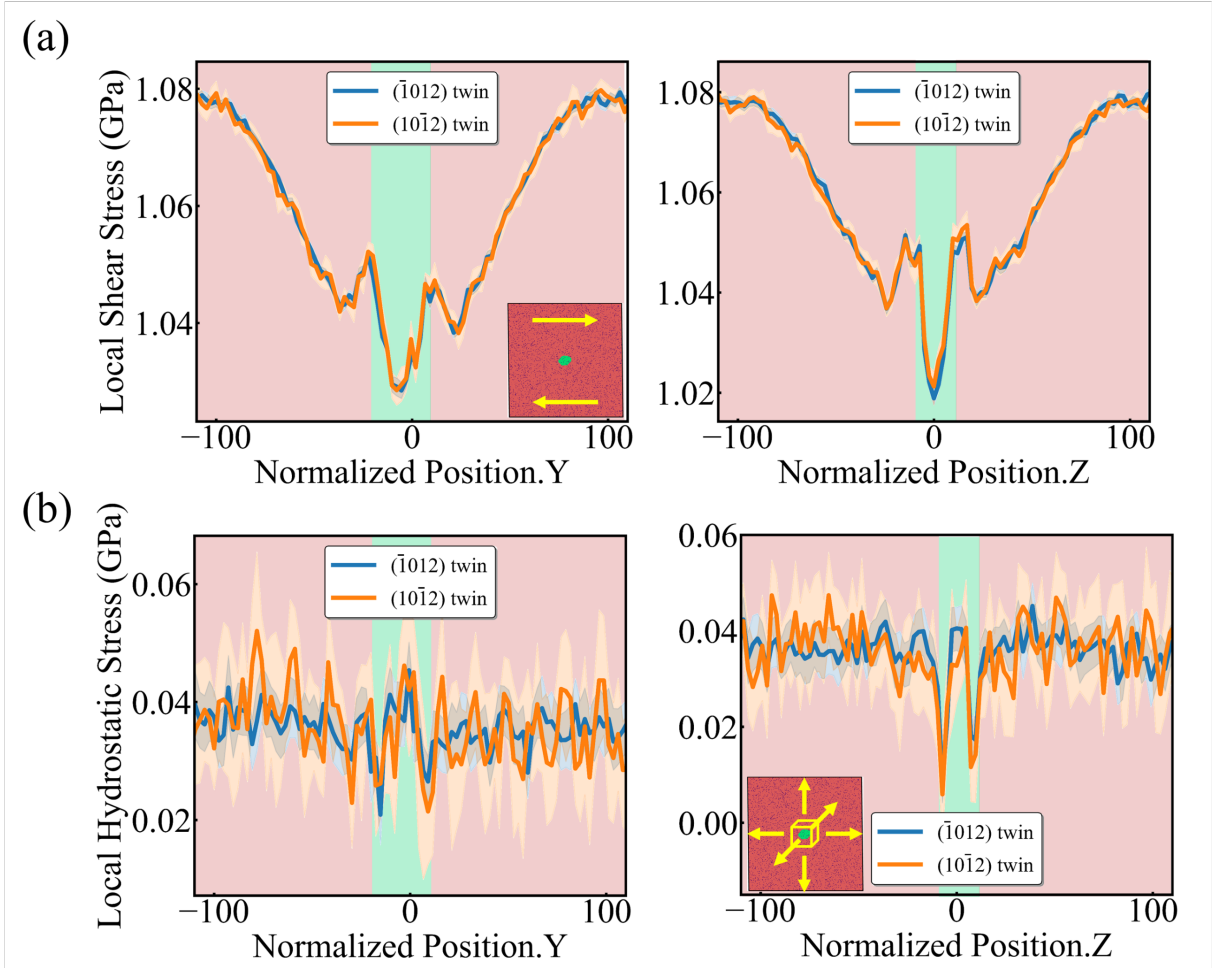

Fig. S8 The distribution of initial (a) shear stress and (b) hydrostatic stress along the Y-axis and Z-axis in samples of Mg-7 at.% Al. The stress values at each Y position or Z position are averaged over samples adopting the same twin variants. The standard deviations are also shown. The insets in (a) and (b) show the directions of shear stress and hydrostatic stress. The light green region in each frame is the twinned region, while the light red region is the matrix.

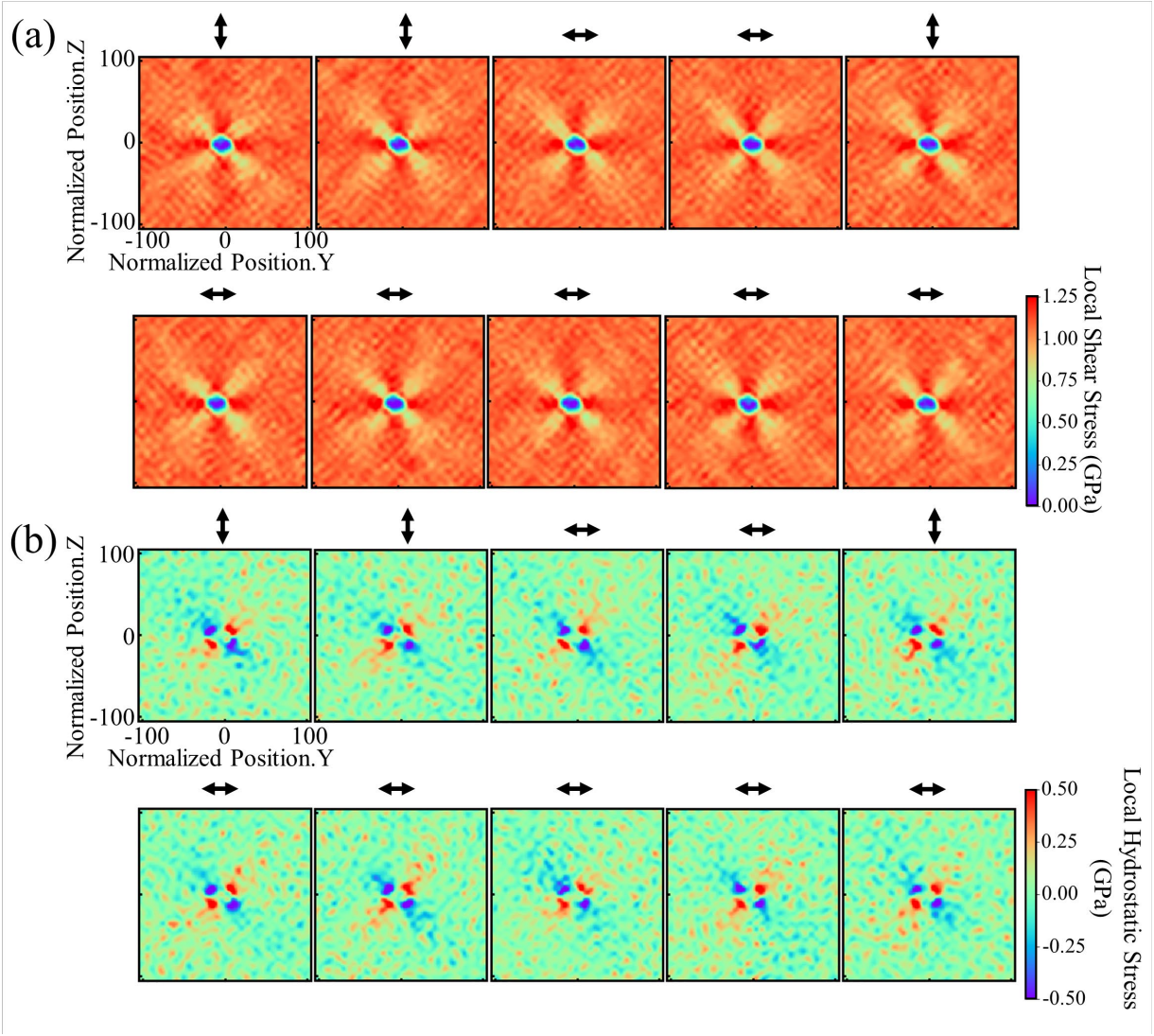

Fig. S9 The 2D distribution of initial (a) shear stress and (b) hydrostatic stress for ten samples of Mg-7 at.% Al. The horizontal and vertical axis show the position (normalized by the lattice constant of Mg-7 at.% Al) along the Y and Z direction of the simulation cell, respectively. The growth direction for twin embryos in each sample is also shown using black arrows. "↔" indicates the $(\bar{1}012)$ twin, while "↕" indicates the $(10\bar{1}2)$ twin.

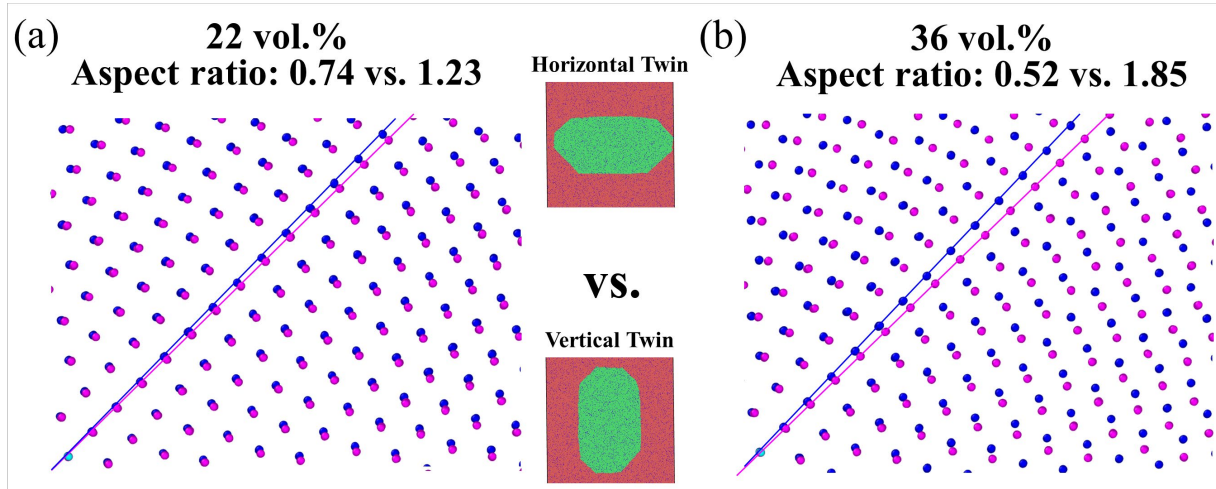

Fig. S10 The atomic positions of twin embryos in two samples with different twinning behavior at (a) 22 vol.% and (b) 36 vol.%. Atoms in the two different twin embryos are colored in magenta and blue, respectively. The basal planes in each twin embryo are marked using magenta and blue lines. The twin embryo configuration at 22 vol.% in each sample are also shown. Although the growth of twin embryos from 22 vol.% to 36 vol.% is strongly influenced by the periodic image, we show (b) here as it is consistent with our statement that as the TBs or c-TBs becomes more dominant in the boundary of twin embryos, the misorientation between basal planes becomes larger.

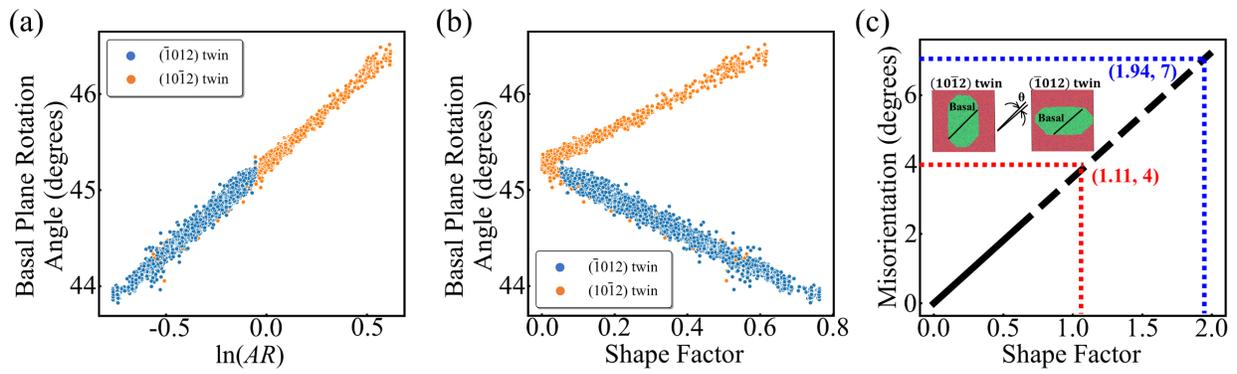

Fig. S11 (a) The same data in Fig. 2(c), but colored using different twin variants. (b) Basal plane rotation angle versus the shape factor of twin embryo for Mg-7 at.% Al samples. (c) The misorientation between basal planes in the two different twin variants versus the shape factor. An estimation of the shape factor at 4° and 7° misorientation is shown.

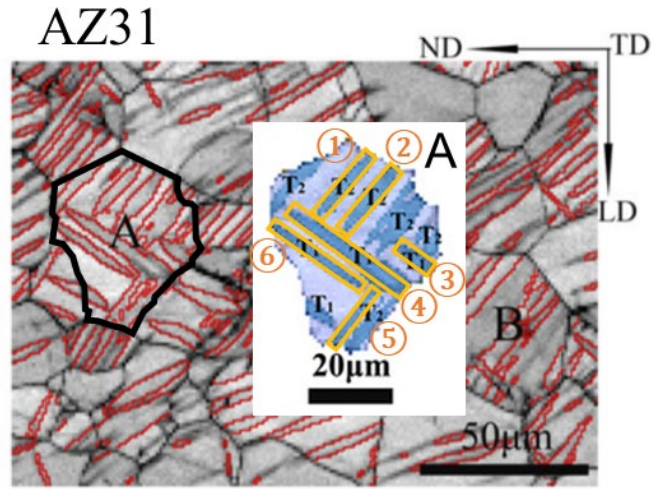

Fig. S12 An Electron Backscatter Diffraction image of the twin morphology in Mg-Al-Zn alloy, with a magnified view of a grain containing co-zone twins shown. The image is reproduced with permission from Ref. [28]. The average aspect ratio and standard deviation is shown below the figure.

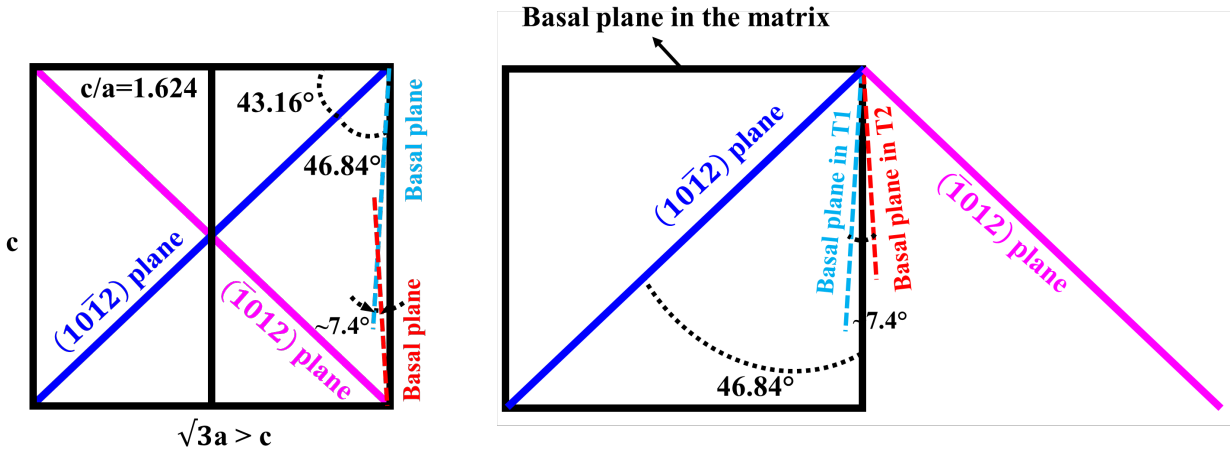

Fig. S13 Schematic showing the deviation in basal plane orientation for co-zone twins. In both figures, the rectangle with length of $\sqrt{3}a$ and height of $c$ is used to represent Mg unit cell viewed from the $[1\bar{2}10]$-direction. The upper and lower side of the rectangle show the orientation of basal planes in the matrix. The magenta and dark blue lines show the twin planes of two co-zone twins. The red dashed line and light blue dashed lines show the basal plane orientation in the two co-zone twins. The misorientation between the twin plane and basal plane is about 43.16°, and the misorientation between basal planes in co-zone twins is about 7.4°.

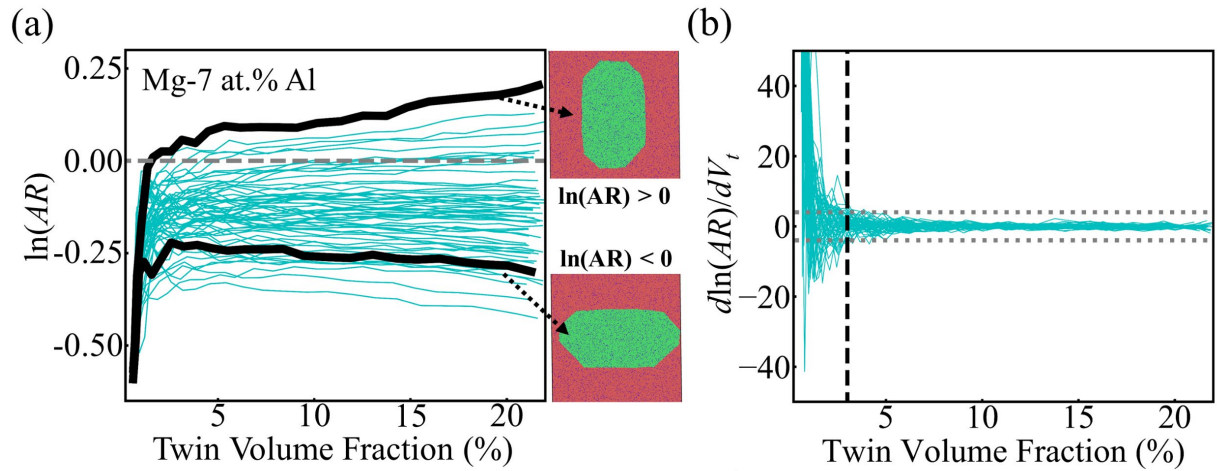

Fig. S14 (a) The variation of $\ln(AR)$ with twin volume fraction for Mg-7 at.% Al samples. The data for two samples are manifested using large black circles, and the twin embryo configuration at 22 vol.% in each sample is also shown. (b) The variation of $d\ln(AR)/dV_t$ with twin volume fraction for Mg-7 at.% Al samples. Grey dotted lines mark the value of ±4, and the black dashed line gives an estimation of the boundary between the incubation period and the deterministic growth stage.

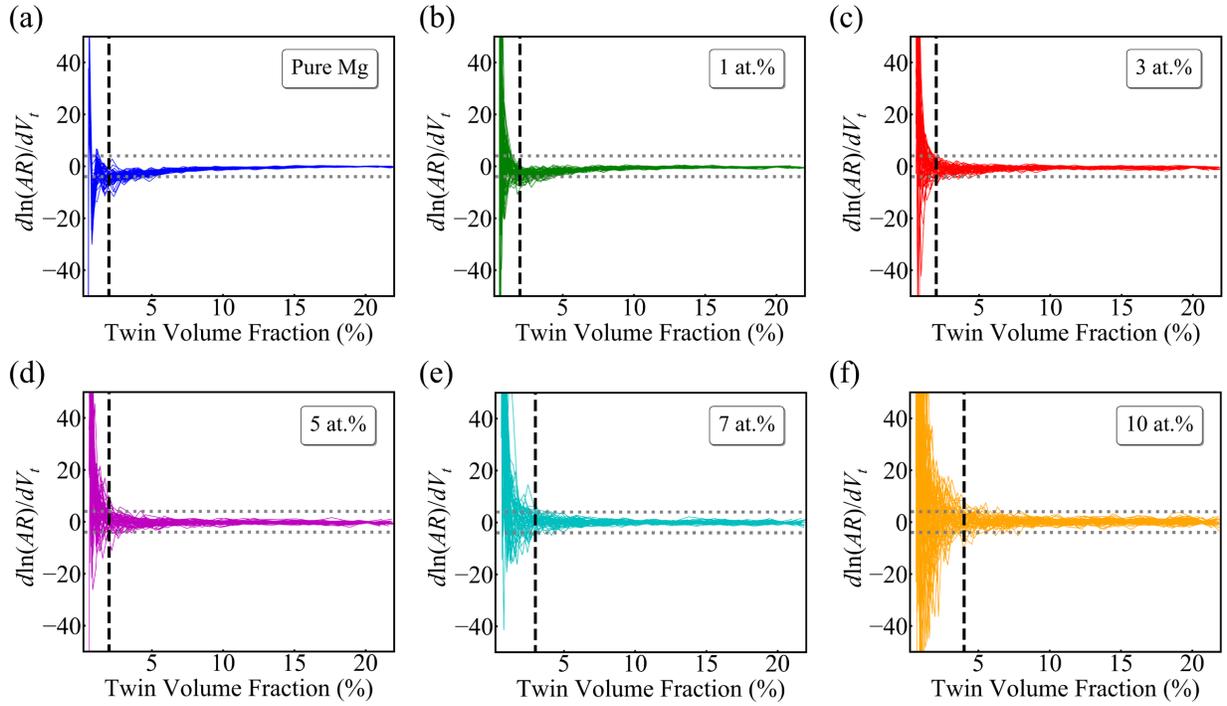

Fig. S15 The variation of $d\ln(AR)/dV_t$ with twin volume fraction for (a) pure Mg, (b) Mg-1 at.% Al, (c) Mg-3 at.% Al, (d) Mg-5 at.% Al, (e) Mg-7 at.% Al, and (f) Mg-10 at.% Al. Grey dotted lines mark the value of ±4, and the black dashed line gives an estimation of the boundary between the incubation period and the deterministic growth stage.

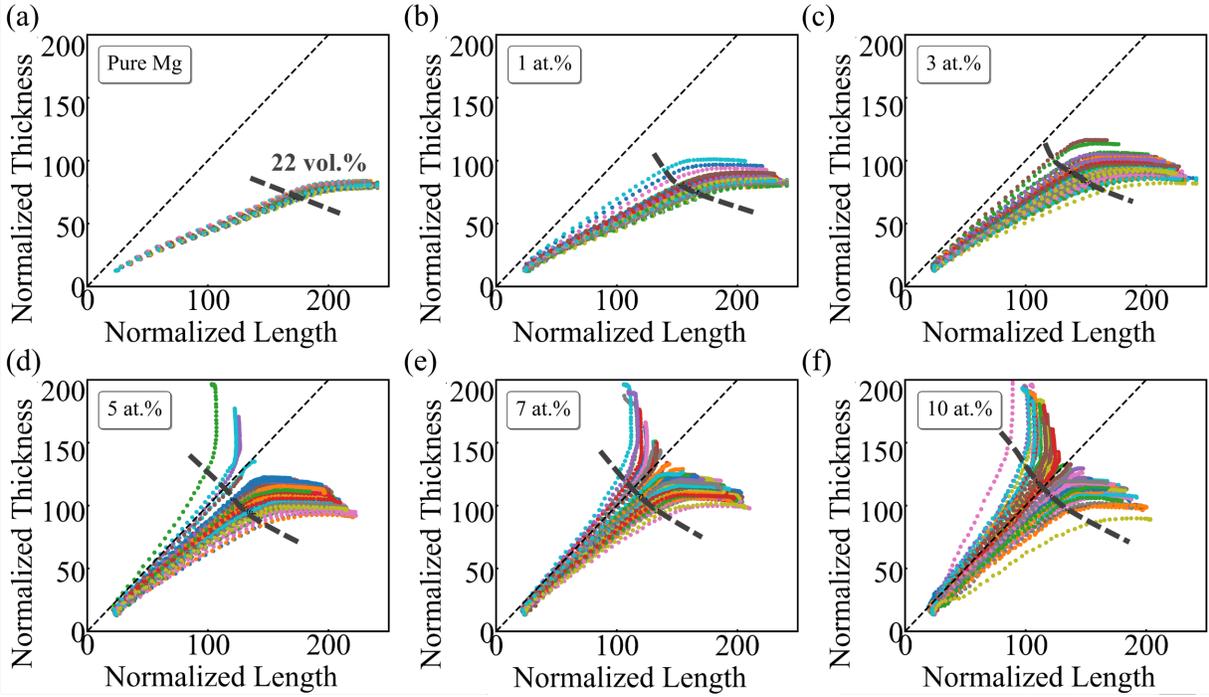

Fig. S16 Raw data of normalized twin thickness versus normalized twin length for (a) pure Mg, (b) Mg-1 at.% Al, (c) Mg-3 at.% Al, (d) Mg-5 at.% Al, (e) Mg-7 at.% Al, and (f) Mg-10 at.% Al. The 1:1 reference lines are shown using black dashed lines. The 22 vol.% lines are also marked as grey dashed lines.

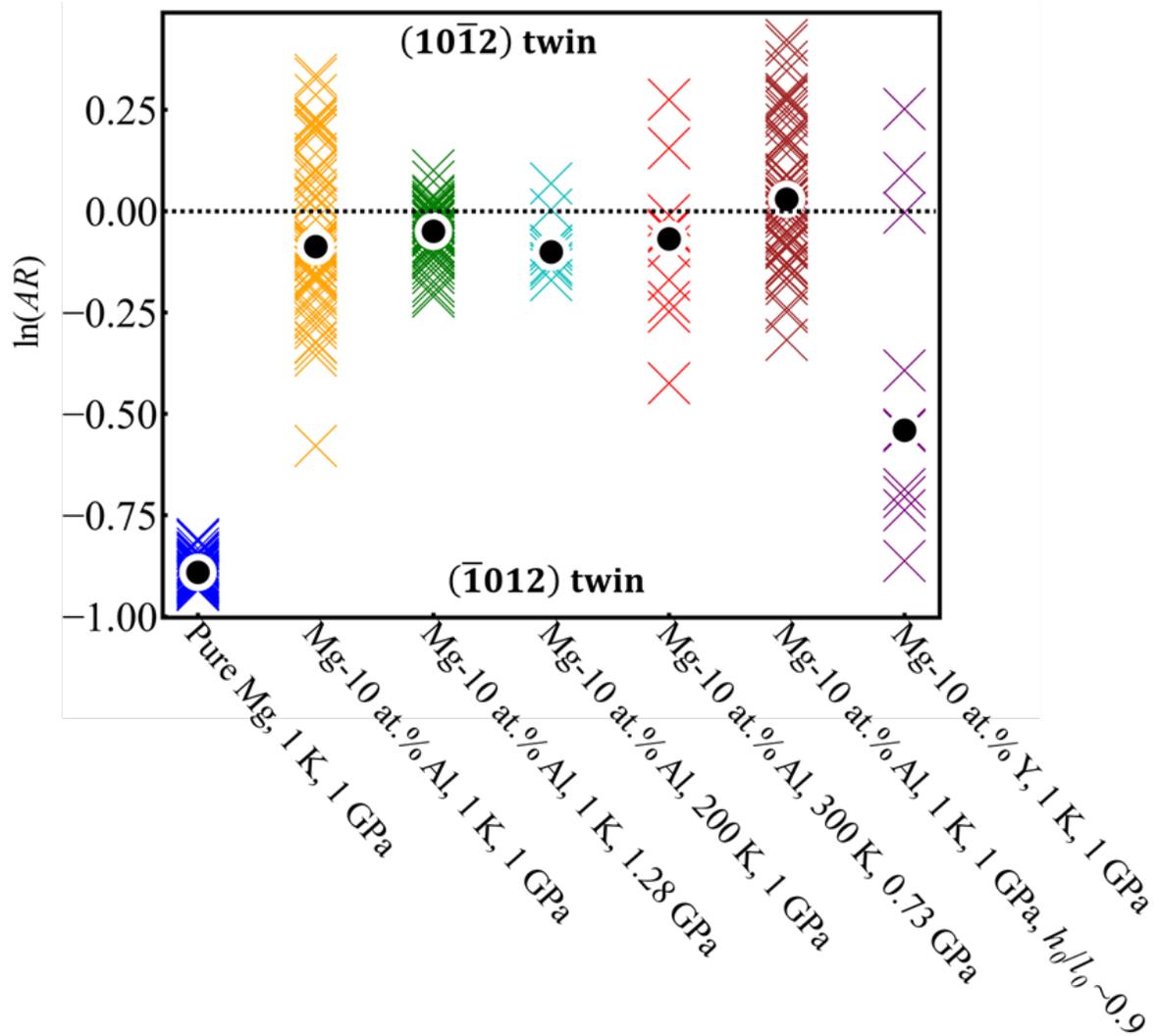

Fig. S17 The natural logarithms of aspect ratio, ln(*AR*), for twin embryos grow at different conditions. All of the alloy samples demonstrate multiple twin variant selection, even as temperature, stress, embryo shape and size, and alloying element are varied.

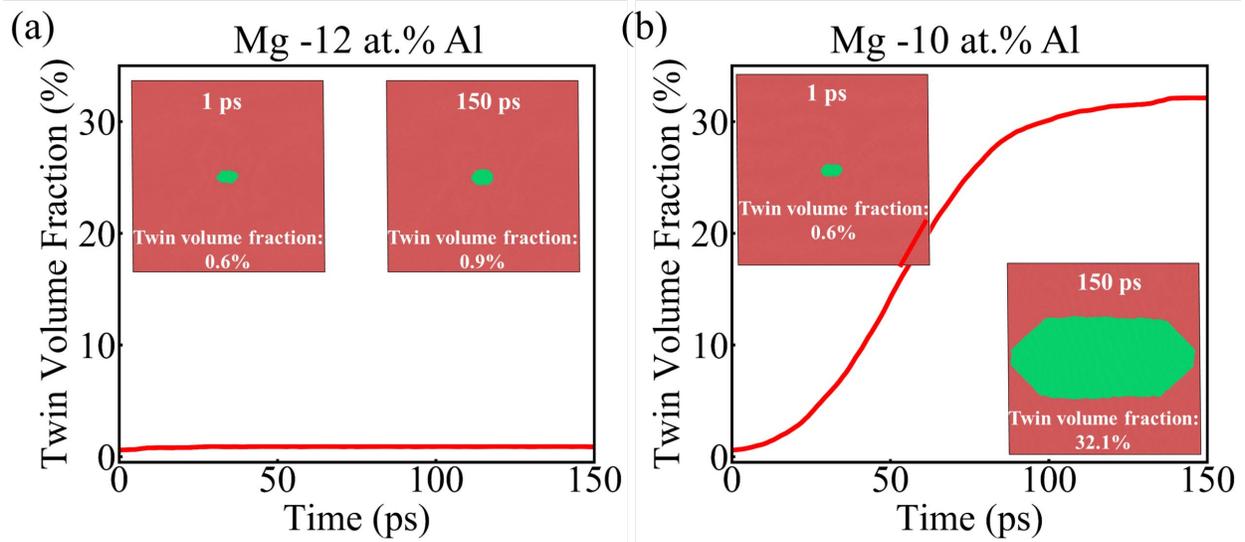

Fig. S18 Twin volume fraction versus time in one sample of (a) Mg-12 at.% Al, and (b) Mg-10 at.% Al. The insets in (a) and (b) show the configuration of twin embryo at 1 ps and 150 ps with the twin volume fraction being revealed.

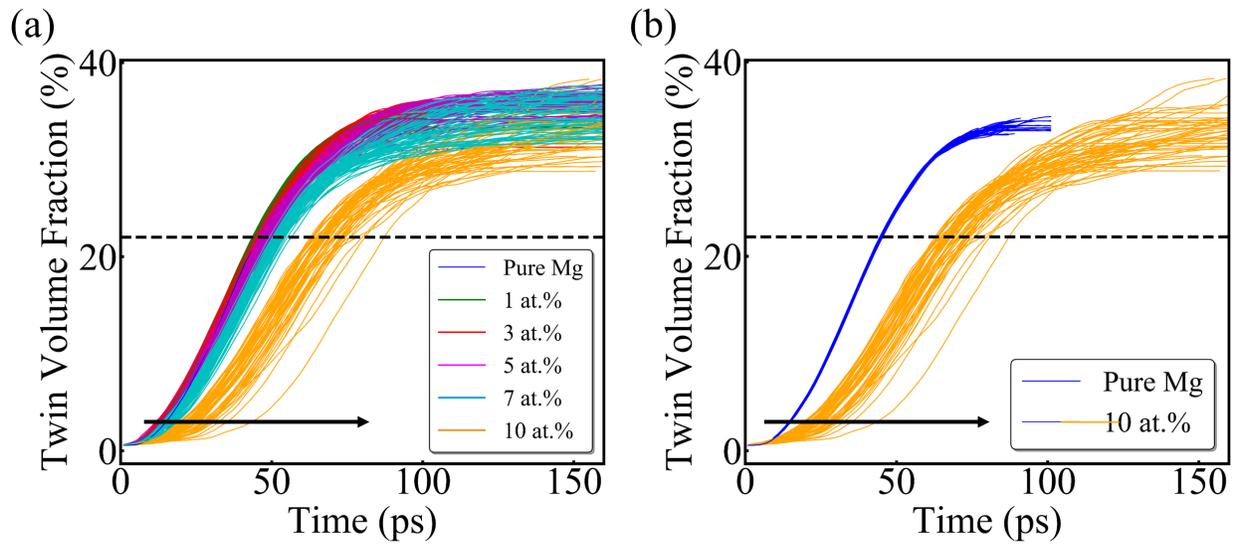

Fig. S19 The twin volume fraction versus time for (a) pure Mg and Mg-Al alloys with all concentrations, and (b) pure Mg and Mg-10 at.% Al alloys. As the Al concentration increases, curves of twin volume fraction versus time start to shift to the right, meaning slower growth of twin embryos.

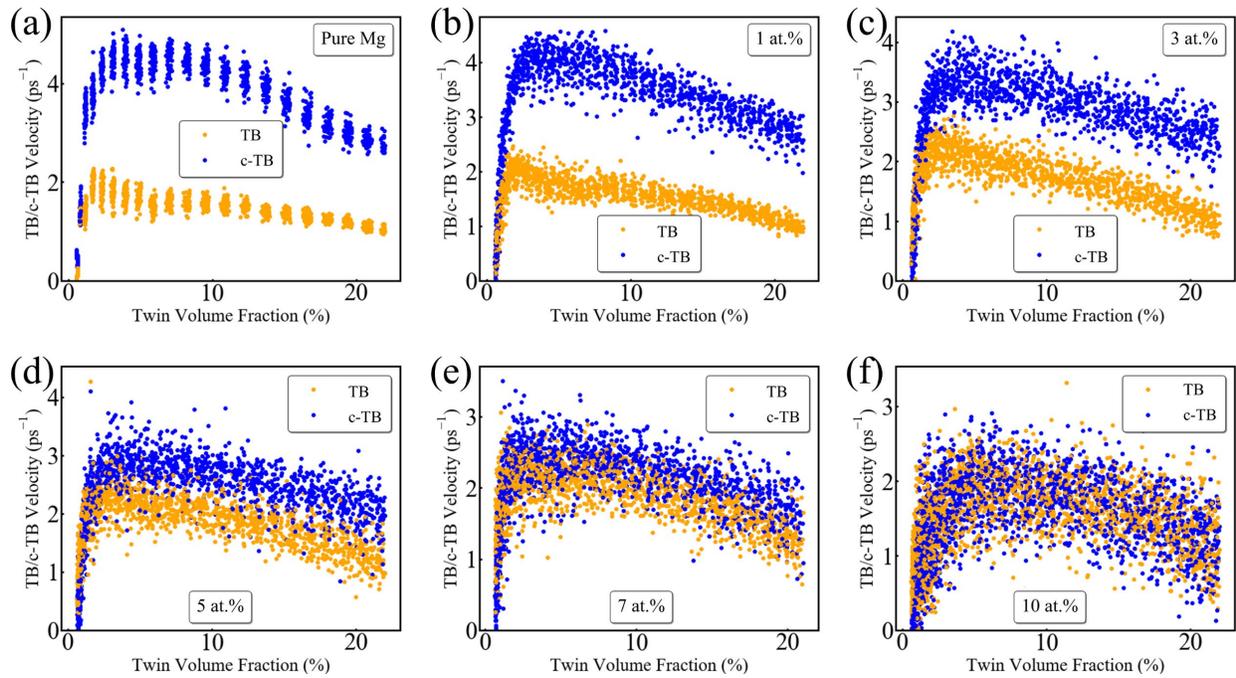

Fig. S20 TB and c-TB velocity versus twin volume fraction, compiled for all samples, for (a) pure Mg, (b) Mg-1 at.% Al, (c) Mg-3 at.% Al, (d) Mg-5 at.% Al, (e) Mg-7 at.% Al, and (f) Mg-10 at.% Al. Alloying alters the magnitude of the TB and c-TB migration, resulting in overlapping velocity curves as composition is increased. These overlapping curves signify the beginning of multiple twin variant selection.

(a) Pure Mg

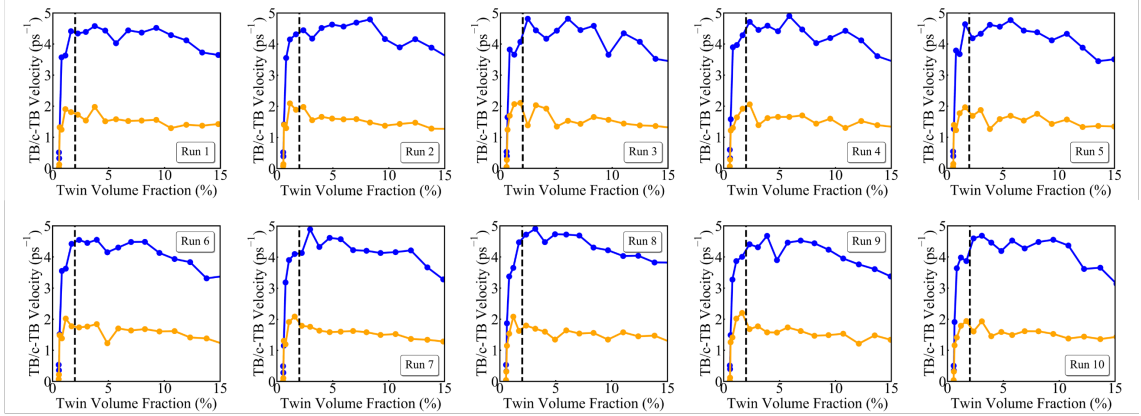

(b) Mg-10 at.% Al

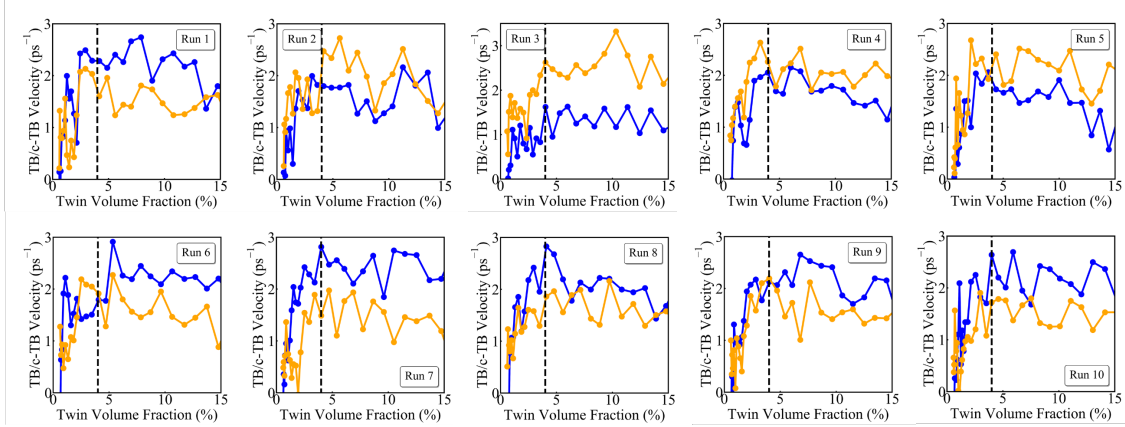

Fig. S21 TB and c-TB velocity versus twin volume fraction for ten individual samples of (a) pure Mg and (b) Mg-10 at.% Al.

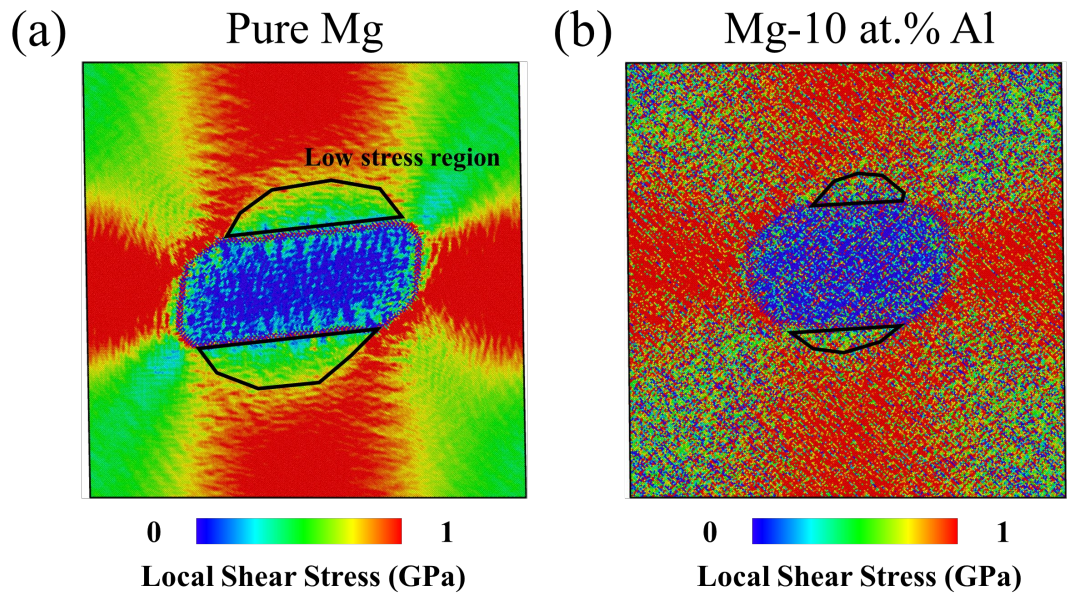

Fig. S22 The local shear stress in one sample of (a) pure Mg, (b) Mg-10 at.% Al. The low shear stress regions are marked using solid black lines.

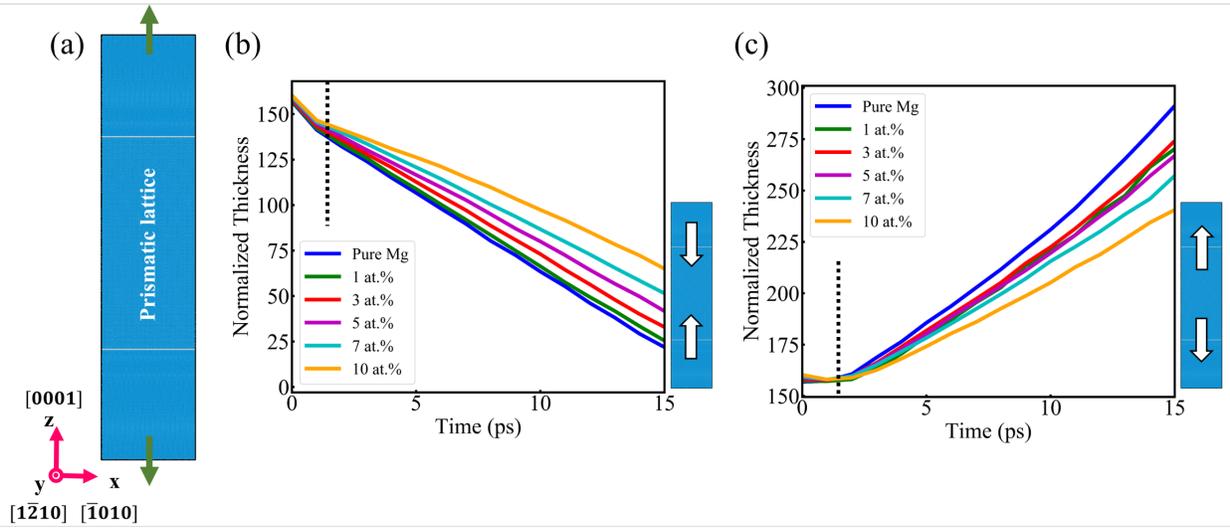

Fig. S23 (a) A schematic of the bi-crystalline sample with two BP/PB interfaces. (b) The normalized thickness of the prismatic lattice region versus time under compressive stress. (c) The normalized thickness of the prismatic lattice region versus time under tensile stress. The black dashed lines in (b) and (c) show after what time the stress fluctuation around 2 GPa becomes moderate.

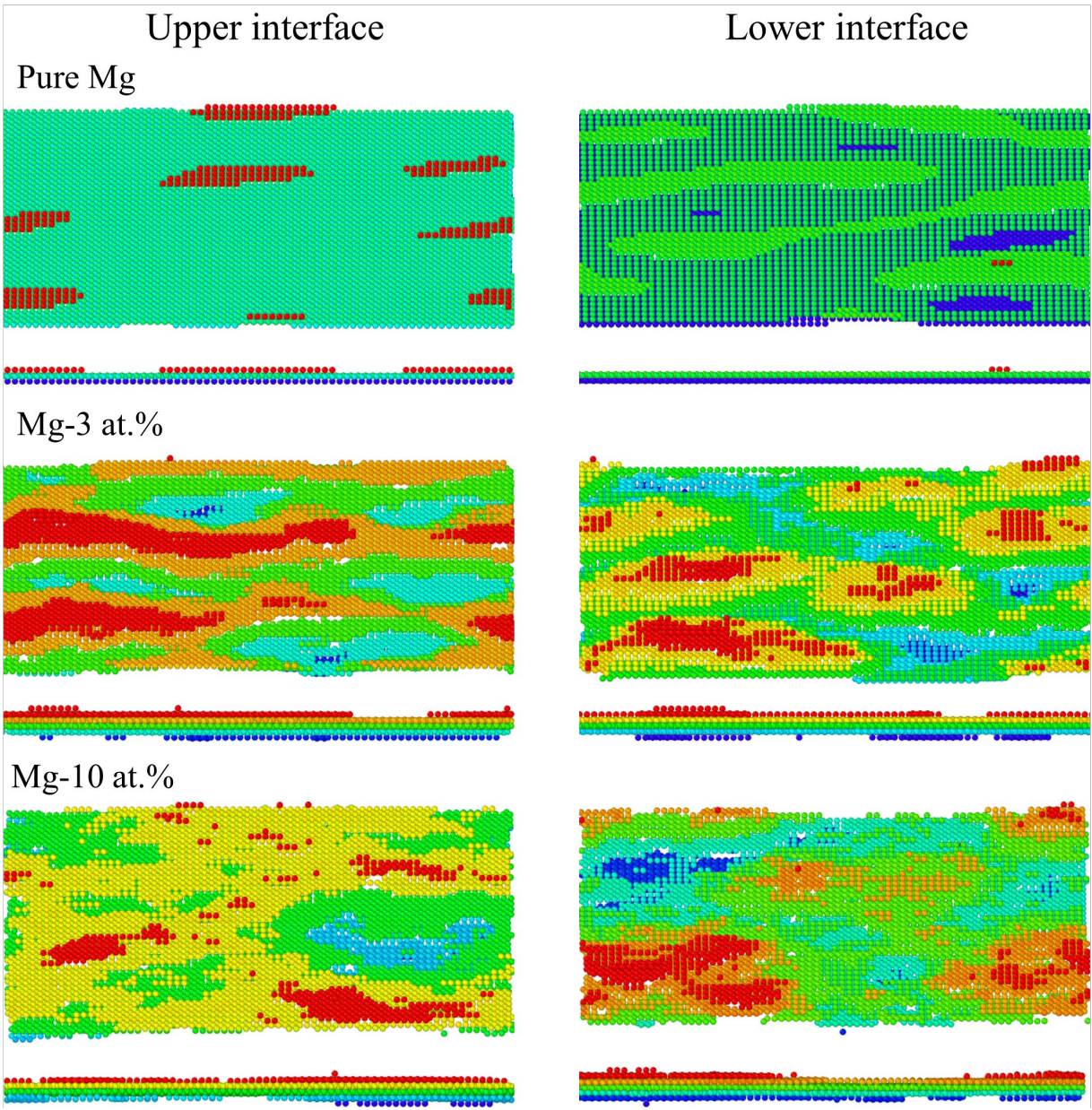

Fig. S24 The perspective view and side view of the upper and lower BP/PB interfaces for pure Mg, Mg-3 at.% Al and Mg-10 at.% Al. Atoms are colored using their Z positions, with red color for atoms on the top and dark blue color for atoms at the bottom.